\documentclass[review]{elsarticle}

\usepackage{hyperref}

\journal{Journal of Systems and Software}

\usepackage[T1]{fontenc}
\usepackage[utf8]{inputenc}

\usepackage{amsmath,amssymb,amsfonts}
\usepackage{graphicx,tikz}
\usepackage{textcomp}
\usepackage{xcolor}
\usepackage{booktabs,tabularx}
\usepackage{listings}
\usepackage{url}
\usepackage{float}
\usepackage{siunitx}
\usepackage{pdfpages}
\usepackage{setspace}
\usepackage{csquotes}

\usepackage{microtype}
\usepackage{enumitem}

\usetikzlibrary{positioning,shapes,arrows,arrows.meta}

\usepackage{ifthen}
\usepackage{amssymb}
\usepackage[normalem]{ulem}

\def\BibTeX{{\rm B\kern-.05em{\sc i\kern-.025em b}\kern-.08em
    T\kern-.1667em\lower.7ex\hbox{E}\kern-.125emX}}
\def\TextIris{\textsc{Iris}}

\usepackage{listings}

\definecolor{colDelimiter}{HTML}{444444}
\definecolor{colKeyword}{HTML}{006600}
\definecolor{colNumber}{HTML}{440088}
\definecolor{colEnum}{HTML}{666633}
\definecolor{colComment}{HTML}{888888}

\lstdefinelanguage{lIris}{
    morestring=[b]", morestring=[b]',
    alsoother={.-},
    sensitive=true,
    morekeywords = {if,then,else},
    morekeywords = [2]{T},
    morekeywords = [3]{SUM,PRODUCT,AND,OR,MIN,MAX,UNION,linear,min,max,sin,cos,exp,ln,log,sqrt,num,months,years},
    morekeywords = [4]{PI,inf,true,false},
    morekeywords = [5]{},
}

\lstdefinestyle{jss2020}{%
  frame           = single,
  framerule       = 1pt,
  rulecolor       = \color{gray},
  columns         = flexible,
  stringstyle     = {\color{colEnum}},
  commentstyle    = \color{colComment},
  keywordstyle    = {\color{colKeyword}\bfseries},
  keywordstyle    = [2]{\color{colKeyword}},
  keywordstyle    = [3]{},
  keywordstyle    = [4]{},
  keywordstyle    = [5]{\color{delimiter}},
  basicstyle      = \ttfamily,
  numbers         = none,
  numberstyle     = \sffamily\scriptsize\color{gray},
  breaklines      = true,
  extendedchars   = true,
  postbreak       = {\hbox{\textcolor{lightgray}{$\hookrightarrow$} }},
  escapeinside    = {(@tex:}{@)},
  keepspaces      = false,%
  breakatwhitespace%
}

\def\jssnumberstyle#1{\textcolor{colNumber}{#1}}

\makeatletter
\newif\ifjss@lst@lastwas@character@
\newif\ifjss@lst@lastwas@charkeep@

\def\jss@@lst@hl@digit#1{\begingroup%
\ifjss@lst@lastwas@character@%
\global\jss@lst@lastwas@charkeep@true#1\else\global\jss@lst@lastwas@charkeep@false%
  \ifnum\lst@mode=\lst@Pmode\relax%
  {\jssnumberstyle{#1}}%
  \else#1\fi\hbox{}%
\fi%
\endgroup%
}
\let\jss@dgt\jss@@lst@hl@digit%

\def\soldisablenumhl{\def\jss@dgt##1{##1}}
\def\solenablenumhl{\let\jss@dgt\jss@@lst@hl@digit}
\def\lst@ProcessLetter{%
  \lst@whitespacefalse\jss@lst@lastwas@character@true\lst@AppendLetter%
}%
\def\lst@ProcessOther{%
  \lst@whitespacefalse\jss@lst@lastwas@charkeep@false\jss@lst@lastwas@character@false\lst@AppendOther%
}%
\def\lst@whitespacetrue{\ifjss@lst@lastwas@charkeep@\jss@lst@lastwas@charkeep@false\jss@lst@lastwas@character@false\else\jss@lst@lastwas@character@false\fi\global\let\lst@ifwhitespace\iftrue}%
\def\lst@whitespacefalse{\global\let\lst@ifwhitespace\iffalse}%

\def\@nc{\ifjss@lst@lastwas@charkeep@\else\jss@lst@lastwas@character@false\fi}

\def\lst@Literate#1#2#3{%
\ifx\relax#2\@empty\else%
\lst@CArgX #1\relax\lst@CDef%
  {}%
  {\let\lst@next\@empty%
  \lst@ifxliterate%
  \lst@ifmode \let\lst@next\lst@CArgEmpty \fi%
  \fi
  \ifx\lst@next\@empty%
    \ifx\lst@OutputBox\@gobble\else%
    \lst@XPrintToken \let\lst@scanmode\lst@scan@m%
    \lst@token{#2}\lst@length#3\relax%
    \lst@XPrintToken%
    \lst@whitespacefalse\@nc%
    \fi
    \let\lst@next\lst@CArgEmptyGobble%
  \fi
  \lst@next}%
  \@empty%
\expandafter\lst@Literate%
\fi}

\makeatother

\lstnewenvironment{iris}[1][]{\lstset{style=jss2020,language=lIris,#1}}{}
\newcommand*\irisinline[2][]{\lstinline[style=jss2020,language=lIris,keepspaces=true,postbreak={},#1]`#2`}

\usepackage{url}
\expandafter\def\expandafter\UrlBreaks\expandafter{\UrlBreaks\do\/\do-}

\usepackage{lineno}
\modulolinenumbers[5]

\newcommand*\patchAmsMathEnvironmentForLineno[1]{%
  \expandafter\let\csname old#1\expandafter\endcsname\csname #1\endcsname
  \expandafter\let\csname oldend#1\expandafter\endcsname\csname end#1\endcsname
  \renewenvironment{#1}%
     {\linenomath\csname old#1\endcsname}%
     {\csname oldend#1\endcsname\endlinenomath}}%
\newcommand*\patchBothAmsMathEnvironmentsForLineno[1]{%
  \patchAmsMathEnvironmentForLineno{#1}%
  \patchAmsMathEnvironmentForLineno{#1*}}%
\AtBeginDocument{%
\patchBothAmsMathEnvironmentsForLineno{equation}%
\patchBothAmsMathEnvironmentsForLineno{align}%
\patchBothAmsMathEnvironmentsForLineno{flalign}%
\patchBothAmsMathEnvironmentsForLineno{alignat}%
\patchBothAmsMathEnvironmentsForLineno{gather}%
\patchBothAmsMathEnvironmentsForLineno{multline}%
}

\allowdisplaybreaks[2]

\makeatletter
\def\fps@figure{thp}\let\fps@table\fps@figure
\expandafter\g@addto@macro\csname end@float\endcsname{%
\ifnum\@floatpenalty<0 \ifnum\@floatpenalty<-\@Mii
\else\ifhmode \if@ignore \penalty\@M \hskip\z@skip \fi\fi
\fi\fi}
\makeatother

\def\ourtitle{A Domain-Specific Language for Modeling and Analyzing Solution Spaces for Technology Roadmapping}

\makeatletter
\AtBeginDocument{\expandafter\gdef\csname Hy@title\endcsname{\ourtitle}}
\makeatother

\begin{document}
\sisetup{detect-all = true}

\begin{frontmatter}
\title{\ourtitle}

\author{Alexander Breckel, Jakob Pietron, Katharina Juhnke, Florian Sihler, \\ and Matthias Tichy}
\address{Institute of Software Engineering and Programming Languages \\ Ulm University \\ Ulm, Germany \\ firstname.lastname@uni-ulm.de}

\begin{abstract}
The introduction of major innovations in industry requires a collaboration across the whole value chain. A common way to organize such a collaboration is the use of technology roadmaps, which act as an industry-wide long-term planning tool. Technology roadmaps are used to identify industry needs, estimate the availability of technological solutions, and identify the need for innovation in the future.
Roadmaps are inherently both time-dependent and based on uncertain values, i.e., properties and structural components can change over time. Furthermore, roadmaps have to reason about alternative solutions as well as their key performance indicators. Current approaches for model-based engineering do not inherently support these aspects.

We present a novel model-based approach treating those aspects as first-class citizens. To address the problem of missing support for \textit{time} in the context of roadmap modeling, we introduce the concepts of a \textit{common global time}, \textit{time-dependent properties}, and \textit{time-dependent availability}. This includes requirements, properties, and the structure of the model or its components as well. Furthermore, we support the specification and analysis  of key performance indicators for alternative solutions. These concepts result in a continuous range of various valid models over time instead of a single valid model at a certain point of time. We present a graphical user interface to enable the user to efficiently create and analyze those models.
We further show the semantics of the resulting model by a translation into a set of global constraints as well as how we solve the resulting constraint system.
We report on the evaluation of these concepts and the \TextIris\ tool with domain experts from different companies in the automotive value chain based on the industrial case of a smart sensing electrical fuse.
\end{abstract}


\end{frontmatter}

\section{Introduction}

Innovation is a driving force behind the past and future development of society as it results in new products, new services, new production methods, new organizations, etc.~(cf.~\cite{EdisonAT13}). One can distinguish between:
\begin{itemize}
    \item disruptive or breakthrough innovation, which can alter the economic landscape,
    \item sustaining innovation, which focuses on continuously improving existing capabilities,
    \item and basic research, which lays the foundation for innovation (cf.~\cite{Satell2017}).
\end{itemize}
For sustaining innovation of individual products, product families, or even whole industry sectors, technology roadmaps are an established tool to systematically identify innovations, their influencing factors (as well as their dependencies) and to represent them over time~\cite{Kerr.2020,Park.2020}. 
Furthermore, technology roadmaps are used to support the strategic management of technologies, to define needs, predict availability of new and/or evolved technology, compare current technologies with emerging technologies, identify dependencies, and estimate possibilities in order to derive a coherent picture of the future innovation. 
In this context, technology roadmaps have a long tradition (the first formal work dates back to 1987~\cite{Willyard.1987}). They play an important role, for example, in the semiconductor industry, which can be seen for example in the initiatives regarding the \enquote{National Technology Roadmap for Semiconductors} (NACS), \enquote{International Technology Roadmap for Semiconductors} (ITRS) or new efforts regarding the \enquote{International Roadmap for Devices and Systems} (IRDS)~\cite{Kerr.2020}.

In this article, we consider technology roadmaps of the second generation (i.e., \enquote{Emerging Technology Roadmaps}), whose focus, according to the classification of Letaba et al.~\cite{Letaba.2015}, is particularly on predicting the development and commercialization of new emerging technologies and comparing them to current technologies. 

The creation of technology roadmaps is currently missing dedicated tool support. 
For instance, Vatananen et al.~\cite{Vatananan.2012} already mention challenges and research opportunities for tools supporting market and technology analysis in technology roadmap development. Complementing this, Park et al.~\cite{Park.2020} request that future research investigate the digitalization aspect of roadmapping by using software and web tools to help organizing workshops, capturing and visualizing data graphically. 
Nevertheless, mostly text documents, drawings, and informal communication are used~\cite{Rinne.2004}, which is also in line with the feedback we receive from our industry partners. Only sometimes calculations are done in generic spreadsheet tools. Most research and existing work in the area of roadmaps (e.g.~\cite{Garcia.1997,Holmes.2006,Lee.2009,Martin.2012}), focuses on the technology roadmapping process or the process for updating and reviewing technology roadmaps and not on how its content is defined. 

We believe that Domain-Specific Languages (DSLs)~\cite{Deursen.2000,Fowler.2011} for modeling of the above mentioned types of roadmaps can significantly improve the creation, update, review, and communication of roadmaps as well as their quality. Furthermore, a DSL for roadmaps needs to be accompanied by a supporting tool that provides analysis capabilities. Since the target group of people creating roadmaps are non-modeling experts, the accompanying tool has to specifically support the experts of the roadmap domain.

The contribution of this paper is a modeling language and an accompanying tool with a domain-specific user interface, which supports the creation and analysis of technology roadmaps and assists roadmap engineers in technology selection and communication of technology-related decisions.
The language is inspired by SysML~\cite{OMGSysML15}, yet implemented as a standalone domain-specific language in order to provide a more streamlined user experience. It supports the definition of an abstract architecture and extends it with an expression language which supports the formal specification of properties and requirements. The expression language specifically supports uncertainty and roadmap time as first-class citizens, e.g., a required technology is available in June 2027, processor performance is expected to increase according to a defined formula 
in the next 10 years, or legal requirements on fuel efficiency change at 3 different points in time in the future. 
Our approach does not only support modeling but also the analysis of the model.
These analyses consist in solving all mathematical expressions for properties and requirements within the model while taking the dependencies between the hierarchical structure of the component model as well as the use of properties in multiple expressions into account. The results show value ranges of properties, the satisfiability of requirements, the temporal availability of modeled components, and which technological alternatives are chosen based on key performance indicators. Hence, these analyses enable the definition of trustworthy roadmaps which are based on complex mathematical estimations instead of ad-hoc rule-of-thumb estimations.

This is supported by a visualization of analysis results. The development of our modeling language and our tool implementation, called \TextIris: Interactive Roadmapping of Innovative Systems, is based on multiple demonstrators provided by our industrial partners from the automotive domain. A detailed description of the tool and, particularly, its interactive and collaboration features are not part of this paper as we focus on the language, its analysis and the visualization of the analysis results.

Therefore, we believe that our research is an important contribution in this area, as we digitalize an important aspect~\cite{Vatananan.2012} of technological roadmapping as requested by Park et al.~\cite{Park.2020} in the sense of a model-based approach to formally describe and analyze aspects of technology roadmaps, taking into account and involving various stakeholders along the value chain and thereby supporting communication between them. Our approach is thus clearly complementary and supportive to previous approaches and tools (e.g., see listing of tools in~\cite{Vatananan.2012}) in this area.

This paper presents a significantly extended version of our modeling language and tool compared to our previously published work~\cite{SEAA2020}. We added the concept of Key Performance Indicators (KPIs) to provide a way for roadmap engineers to define the quality or suitability of competing solution alternatives, see  Section~\ref{ssc:modeling-language}. We also define our approach of using interval arithmetic to deal with uncertain values in Section~\ref{sec:uncertainty}. Furthermore, we explain the approach to solve the global system of equations spanned by properties, requirements, and KPIs using repeated symbolic transformations. Thereafter, we describe in Section~\ref{sec-solver} the three transformation phases and the actual solving process of the constraint system. Finally, we additionally evaluated our modeling language and tool with industrial partners along the value chain and present the results of the case study in Section~\ref{sec:evaluation}. The industrial experts value the modeling language and analysis capabilities and, particularly, that our language enables the various contributions of the partners along the value chain to be captured separately, that  solution spaces created by alternative solutions and KPIs are created and analysed easily and quickly, and that \TextIris\ offers the possibility to evaluate even complex systems and edge solutions.

After an overview of related work in the next section, we introduce our running example in Section~\ref{sec-example}. This running example is a simplified version of a roadmap for an electronic fuse in the automotive domain. We present our modeling language in Section~\ref{sec-concepts} and describe the solver we implemented to analyze time-dependent constraint systems with uncertain values in Section~\ref{sec-solver}. In the following Section~\ref{sec-visu}, we describe the user interface and visualization concepts of \TextIris. In addition, we present the results of the evaluation of our modeling language and the corresponding tool with industrial domain experts in Section~\ref{sec:evaluation} based on the smart sensing fuse presented in Section~\ref{sec-example}. We finish with a conclusion and an outlook on future work in Section~\ref{sec-conclusions}.

\section{Related Work}\label{sec:related-work}

Technology roadmaps are established as a strategic management tool for technology planning, management, and selection~\cite{Rinne.2004,Kerr.2020,Phaal.2004,Kostoff.2001,Vatananan.2012}. They contribute to the planning of technology investment and development and are thus a key element for the success and competitiveness of organisations~\cite{Knoll.2018,Vatananan.2012}. A great number of publications address technology roadmap processes that focus on the development, updating, and maintenance of technology roadmaps~\cite{Garcia.1997,Holmes.2006,Lee.2009,Martin.2012}. 

However, there is a lack of methods and specialized tools for the development and documentation of technology roadmaps. Vatananan et al.~\cite{Vatananan.2012} consider tools as an important component for the development and implementation of roadmaps. They categorize supporting tools for technology roadmap development according to their functionality into (a) market analysis tools, (b) technology analysis tools, and (c) supporting tools. The domain-specific language and the corresponding tool presented in this paper belong to the group of technology analysis tools, as predictions about technologies are mapped over time, based on their key performance indicators (KPIs). 

Nowadays, technology roadmaps are often captured using Microsoft PowerPoint or Visio~\cite{Rinne.2004} and are therefore visualized graphically or even as visual models. According to their graphical format, Phaal et al.~\cite{Phaal.2004} distinguish eight types of technology roadmaps: multiple layers, single layer, bars, tables, graphs, pictorial representations, flow charts, and text. 
A major disadvantage of such technology roadmaps is that their adjustment, maintenance and reuse of individual components is difficult. 
Technical analyses of certain KPIs based on technical models are not part of simple graphical representations, which have been created using Microsoft PowerPoint, for example. This makes verification and validation of technology objectives more difficult~\cite{Knoll.2018}. 

To overcome this, Knoll et al.~\cite{Knoll.2018} present a model-based approach where experts create technical models that enable the evaluation of potential product architectures according to a defined set of KPIs at different time horizons. Therefore, the roadmap architecture is described in an Object Process Methodology (OPM) diagram (cf. ISO~19450~\cite{ISO.19450}). This diagram includes a list of existing products and services, elements of technological relevance, primary operational functions, and roadmap related KPIs. The additional necessary data of the technology roadmap, such as definitions of KPIs, financial models, list of alternative and competitive technologies and their features, technological readiness level for each technology, statistical and mathematical models, are stored in a common database. 
How these information are linked to each other, how they are integrated into a technology roadmap, and how they are visualized is not described in detail by the authors.
Furthermore, they do not describe a modeling language nor which analyses based on the technical models are possible and how they could be performed.

In contrast, our paper contributes to model-based roadmapping and focuses explicitly on visualization and interaction concepts that assist roadmap engineers.
For this purpose, we specifically present the modeling language for the creation of a technology roadmap, which supports roadmap engineers in analyzing underlying technical models and their interrelationships.

Our definition of \emph{time} differs from other usages of time commonly found in formal models, like in Time Petri Nets~\cite{CR05} or in Timed Automata~\cite{Alur&Dill1994}. These formalisms define time-dependent \emph{behaviour} based on real-time clocks, in order to introduce ways to delay and synchronize the execution of models. Other approaches, like real-time extensions of the UML~\cite{OMG2009}, support the specification of real-time behaviour of embedded systems with a focus on performance and schedulability. In contrast to that, our work defines time-dependent \emph{structures} to characterize the future availability of the modeled products and technologies over a time-frame of months, years, or decades.

    Another important issue is uncertainty, as many factors cannot be accurately estimated. Lee et al. address this problem by introducing a Bayesian Belief Network (BBN) that structures multiple possible scenarios~\cite{lee2010bayesian}. BBNs and their use for technology roadmapping are discussed in the context of flexibility and adaption to risk by Jeong et al.~\cite{jeong2021developing}.
    In \TextIris, we use ternary logic (cf.~\cite{kleene1938notation}) and interval arithmetic (cf.~\cite{moore1966interval}) to model uncertainty. This allows for more convenient use with the other constructs of our domain-specific language, as BBNs would introduce another level of complexity. Currently, the interval arithmetic and ternary logic seem to be sufficient. However, if they are identified to be inadequate at some point in the future, we may extend our domain-specific language to include concepts like BBNs. 

    Our work is also related to the area of software and system architecture as our language is strongly based on the idea of hierarchical component models. Architecture Description Languages (ADLs) (cf.~\cite{DBLP:journals/tse/MedvidovicT00}) have been introduced more than two decades ago to define the architecture of systems more explicitly. Our hierarchical component model follows the concepts outlined for architectural description languages by Medvidovic and Tayler~\cite{DBLP:journals/tse/MedvidovicT00}.
    
    The Architecture Analysis and Design Language (AADL) is a prominent example of an architectural description language which can be extended by so-called annexes. Annexes exist for example for time, e.g., \cite{DBLP:journals/sttt/MkaouarZHJ20}. However, those time extensions deal (similar to the aforementioned UML profile~\cite{OMG2009}) with time aspects relevant for real-time scheduling. Bao et al.~present an AADL annex for the specification of uncertainty~\cite{Bao+2017}. While their approach enables modeling of uncertainty w.r.t.~to a diverse set of probability distributions, they also target detailed performance and safety requirements during product development. Furthermore, those approaches~-- in contrast to our approach~-- do not address the specification and selection of alternative solutions based on user-defined KPIs.
    
    The research area of architecture optimization specifically supports the optimization w.r.t.~to user-defined KPIs. Aleti et al.~classify many different architecture optimization approaches in their survey~\cite{DBLP:journals/tse/AletiBGKM13}. However, these approaches focus on optimizing a concrete product during development time and do not allow specifying and analyzing future solution spaces. 
    
    Finally, software product lines~\cite{DBLP:books/daglib/0032924} are related to our approach as they enable the specification of different alternative features. There exist approaches to select features based on objective functions similar to KPIs in our approach, e.g.,~\cite{DBLP:journals/tosem/HieronsLLSZ16}. However, those approaches also focus on concrete products and do not support modeling and analyzing future solution spaces.

\section{Running Example}\label{sec-example}

In order to illustrate the developed concepts and visualizations presented over the course of this paper, we will use a simplified example called \textit{Fuse} as a running example.

Every electrical device in modern cars is protected from overcurrent by so called blade fuses. In case of overcurrent, a small metal strip inside the fuse melts, the electrical circuit is interrupted and the connected device is switched off. To switch the device on again, the fuse has to be physically replaced. Depending on the layout of a car's wiring harness and the particular melted fuse, more than just the actual faulty loads might be switched off in case of an overcurrent.

One major research goal of the automotive industry is to develop autonomously driving vehicles.
In case of an overcurrent, such a vehicle has to be stopped safely.
Since blade fuses can not melt in a controlled way, in a worst-case scenario the central processing unit which controls and stops the car could be switched off as a side-effect of a faulty electric device. In such a case, the car would be out of control.

As an alternative to blade fuses, a reversible smart sensing fuse is not only able to detect and protect against overcurrents but also selectively switch off electrical consumers and switch them back on again without being replaced. If the vehicle's electrical system is adapted accordingly and each consumer is individually protected, in case of an overcurrent, the car could be stopped safely.

Automotive Original Equipment Manufacturers (OEMs) and Tiers (direct or indirect suppliers of the OEM) have been researching semiconductor-based reversible smart sensing fuses for more than two decades~\cite{graf_sense_1996, fuisting_current_2014}. A big challenge was and still is to design a semiconductor switch that supports high currents, has an adequate avalanche rating, and beneficial feature set. The design involves several design decisions and key technologies with varying market availability.

\begin{figure*}[tpb]
    \centering
    \includegraphics[width=\textwidth,height=.95\textheight,keepaspectratio]{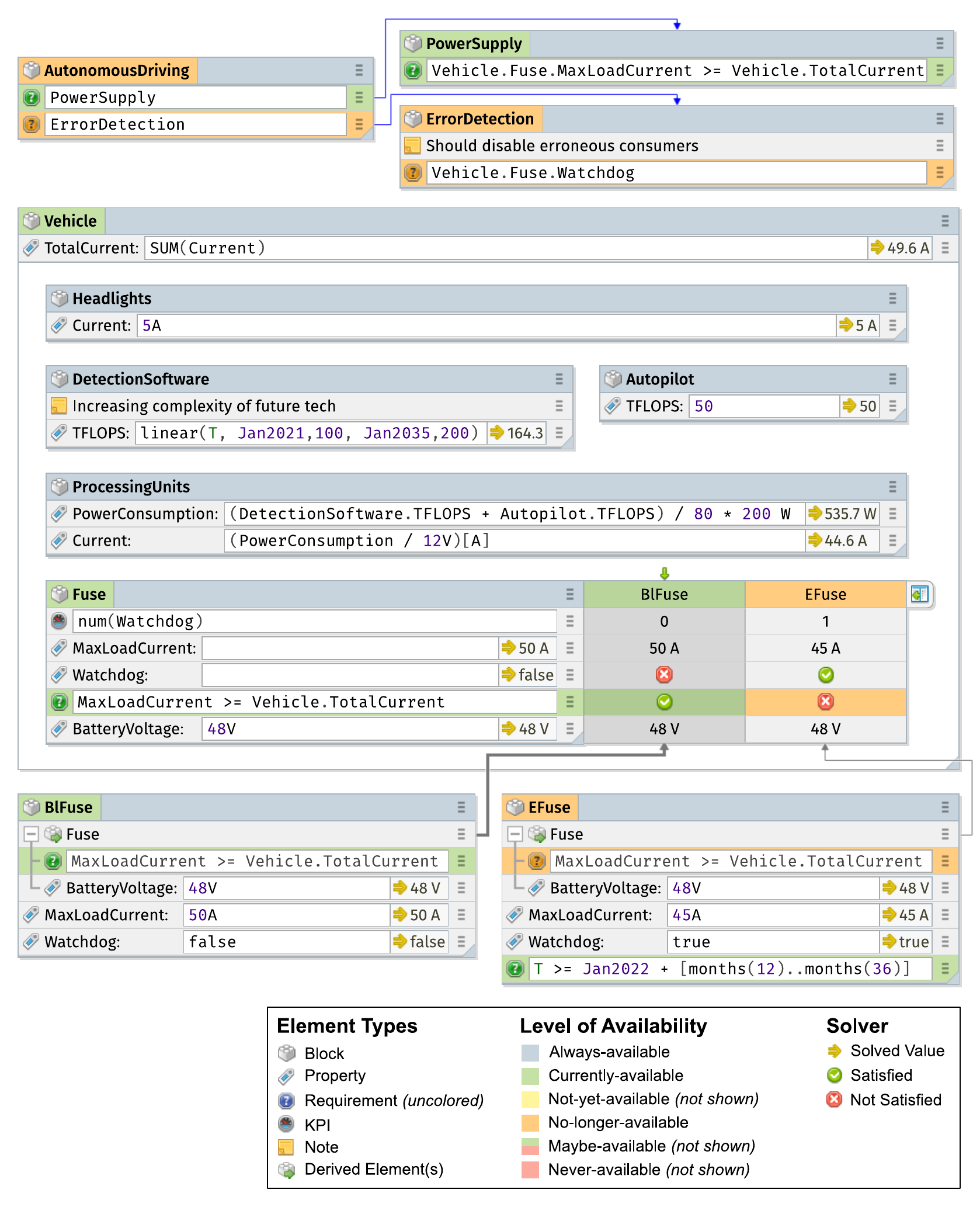}
    \caption{Screenshot of a model of two alternative fuse technologies (BlFuse and EFuse) within the context of an autonomously driving vehicle ($T=Jan 2030$)}
    \label{fig:runingExample}

\end{figure*}

Figure~\ref{fig:runingExample} shows a screenshot of the \textit{Fuse} example modeled with \TextIris. The example is arranged in three horizontal imaginary layers. The layers themselves have no semantic meaning and are for layout purposes only. On the top layer the abstract functionality \texttt{AutonomousDriving}, which should be available in the future, is modeled as a block depending on the availability of two features: the fuse's max load current must be greater or equal than the  path load worst case operating current. Additionally, a selective overcurrent detection and consumer disabling must be available. This dependence in terms of availability is expressed by \textit{requirements} which are a special model elements referencing another block's availability or holding a Boolean expression. There are different levels of availability as illustrated in Figure~\ref{fig:runingExample}.

To evaluate, if and when in the future it is possible to develop an autonomously driving car, in the second layer, a \texttt{Vehicle} is modeled as a generic block and referenced by the availability requirements of \texttt{PowerSupply} and \texttt{ErrorDetection}. The vehicle holds a property \texttt{TotalCurrent} whose value is the sum of all sub-blocks' \texttt{Current}s. The meaning of properties is twofold: on the one hand, they state that something has a property to be considered in the course of the modeling goal. On the other hand, their values are computed by the solver, e.g., \texttt{TotalCurrent}'s value is evaluated to \texttt{\SI{49.6}{\A}}.

There are two electrical consumers: \texttt{Headlights} with a static \texttt{Current} of \texttt{\SI{5}{\A}} and \texttt{ProcessingUnits} responsible for running software that is needed for the autonomous driving of the car. In the example, the processing units are responsible for two types of software: the \texttt{DetectionSoftware} that is able to detect obstacles, road signs, and the environment around the car, and the \texttt{Autopilot} software that drives the autonomous car. Both software parts require a specific amount of processing power in \texttt{TFLOPS}. Since future cars will drive more and more autonomously, the \texttt{DetectionSoftware} will become more and more important and extensive. To address this fact, the \texttt{DetectionSoftware} will require more \texttt{TFLOPS} over time as modeled by a linearly growing value of \SI{100}{TFLOPS} in 2021 up to \SI{200}{TFLOPS} in 2035. In consequence, also the \texttt{ProcessingUnits'} overall \texttt{PowerConsumption} and \texttt{Current} grow accordingly. In our running example, \texttt{PowerConsumption} is a rough estimation based on present GPUs, which offer \SI{80}{TFLOPS} with a power consumption of \SI{200}{W}.

The \texttt{Vehicle} also holds a \texttt{Fuse} which defines a solution space and acts as an interface for concrete fuse implementations. Therefore, the block \texttt{Fuse} defines various properties that a concrete fuse must have: a \texttt{Watchdog} to individually detect overcurrents, a supported \texttt{MaxLoadCurrent}, and a \texttt{BatteryVoltage}. While \texttt{BatteryVoltage} is fixed to \SI{48}{\V}, the remaining properties need to be set by the specific implementation. Besides the properties, \texttt{Fuse} defines a requirement that \texttt{MaxLoadCurrent} must be greater or equal than the \texttt{Vehicle}'s calculated \texttt{TotalCurrent}, which in turn~-- as already mentioned~-- grows over time. 

In the third layer there are two concrete fuses which implement the interface defined by \texttt{Fuse}. An \textit{implementation} expresses an \textit{isA} relationship. Since \texttt{BlFuse} and \texttt{EFuse} implement the interface of \texttt{Fuse}, they derive all properties, constraints, and values (if set) defined by \texttt{Fuse}. \texttt{BlFuse} represents a blade fuse with a high \texttt{MaxLoadCurrent} but without a \texttt{Watchdog}, and \texttt{EFuse} represents a smart sensing fuse and has a \texttt{Watchdog} but a lower \texttt{MaxLoadCurrent}. Both fuse implementations are also displayed next to the \texttt{Fuse} block which enables manual comparison of different implementations of \texttt{Fuse}. \TextIris\ is also capable of automatically selecting the \textit{best} implementation of arbitrary alternative implementations. In this example, a fuse with a watchdog would be seen as the \textit{best} fuse. Accordingly, \texttt{Fuse} contains a \textit{KPI} that references the existence of an implementation's watchdog. However, at the current point in time (\textit{T=Jan2030}) \texttt{BlFuse} is selected as indicated in the \texttt{Fuse} block by the green arrow above \texttt{BlFuse}, because \texttt{EFuse} has a watchdog but violates the requirement of its \texttt{MaxLoadCurrent}, which must be equal or greater than the vehicle's total current.

\section{A Modeling Language for Roadmapping}\label{sec-concepts}

In this section we present our modeling language in four steps:

\begin{enumerate}
    \item First, we introduce a domain-specific modeling language that is capable of representing the structural composition of a technological system relevant for technology roadmapping.
    \item Second, we introduce the embedded textual language used to specify property values and requirement conditions.
    \item Third, we extend both languages to support time-dependent properties and the time-dependent availability of structural elements.
    \item And last, we extend both languages to support interval arithmetic and ternary logic as they are required to model uncertainty, common in real-world applications.
\end{enumerate}

\subsection{Structural Modeling Language}\label{ssc:modeling-language}

Our modeling language was developed in continuous coordination with industry partners and was inspired by SysML block definition and internal block diagrams \cite[pp.~35--43]{OMGSysML15}. It is intended to represent structural, hierarchical models that describe the technologies relevant for a technology roadmap, together with their properties, requirements, and KPIs. We used SysML as inspiration due to its wide-spread use in systems modeling. However, we chose a vastly reduced and slightly adapted set of modeling elements due to our target user group of domain experts with little to no experience with modeling formalisms.
Our goal in designing the language was to define a minimal set of modeling elements that, on the one hand, is capable of representing all concepts relevant for roadmapping, while on the other hand being comprehensible enough to be applied by domain experts. Instead of defining a standalone language, it would have also been possible to extend SysML with KPIs, time-dependence, and solution spaces. However, this would cause our language to inherit several technical aspects of SysML that are not needed due to our reduced feature set, and would be detrimental to the overall user experience. For example, the distinction in SysML between types, occurrences and instances would require model elements to be defined redundantly. Also, the overall usage of parametric diagrams to specify mathematical dependencies would add significant modeling overhead when using our embedded expression language.

An important aspect of using formal models for technology roadmapping is the ability to model solution alternatives for a technological need.
A required functionality, like the protection from overcurrent in an electrical circuit, may be achieved in different ways, e.g., through a blade fuse or a smart sensing fuse.
Therefore, our language supports solution alternatives.

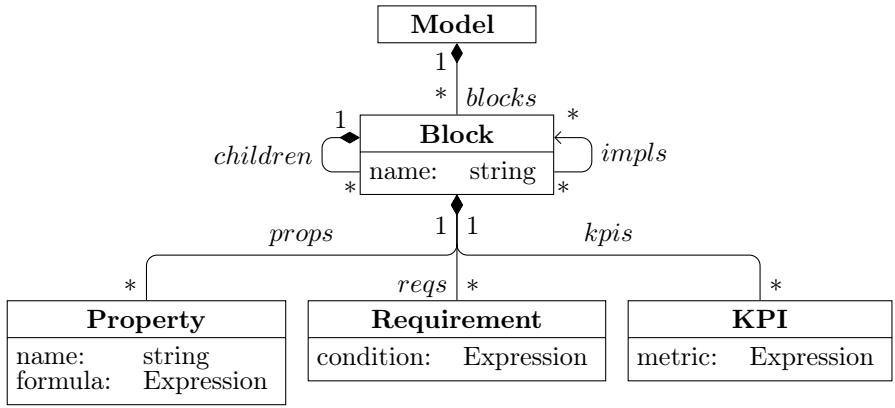
\begin{figure}
    \newenvironment{umlprop}{\tabular{@{}l<{:}l@{}}}{\endtabular}%
    \centering%
    \begin{tikzpicture}[every node/.append style={%
            node font=\fontsize{10}{8}}]
        \tikzstyle{box}=[rectangle, draw=black, anchor=north, align=center, rectangle split, rectangle split parts=2, text depth=0pt]
        \tikzstyle{aggregation}=[->, >=open diamond, rounded corners=4pt]
        \tikzstyle{composition}=[->, >=diamond, rounded corners=4pt]
        \tikzstyle{association}=[-Straight Barb, rounded corners=4pt]

        \node (Model) [box, minimum width=2.1cm, rectangle split parts=1] at (3,1.6) {
            \textbf{Model}
        };
        \node (Block) [box, minimum width=2.1cm] at ([yshift=-1.2cm]Model) {
            \textbf{Block}
            \nodepart[align=left]{second}%
            \begin{umlprop}
                name & string
            \end{umlprop}
        };
        \coordinate (prop-req-kpi) at ([yshift=-1.4cm]Block.south);
        \node (Property) [box, minimum width=3.2cm,below left,xshift=-15ex] at (prop-req-kpi) {
            \textbf{Property}
            \nodepart[align=left]{second}
            \begin{umlprop}
                name    & string \\
                formula & Expression
            \end{umlprop}
        };
        \node (Requirement) [box, minimum width=3.2cm,below] at (prop-req-kpi) {
            \textbf{Requirement}
            \nodepart[align=left]{second}
            \begin{umlprop}
                condition & Expression
            \end{umlprop}
        };
        \node (KPI) [box, minimum width=3.2cm,below right,xshift=15ex] at (prop-req-kpi) {
            \textbf{KPI}
            \nodepart[align=left]{second}
            \begin{umlprop}
                metric & Expression
            \end{umlprop}
        };

        \draw[composition] (Block.190) -- ++(-0.5,0)
            node[near start,below] {*}
            |- (Block.170) node[near start,left] {$children$}
            node[near end,above] {1};
        \draw[association] (Block.350) -- ++(0.5,0)
            node[near start,below] {*}
            |- (Block.10) node[near start,right] {$impls$}
            node[near end,above] {*};
        \draw[composition] (Property.north) -- ++(0,0.6)
            node[near start,left] {*}
            -| node[near start,above] {$props$}
            (Block.south) node[near end,left] {1};
        \draw[composition] (Requirement.north) -- ++(0,0.6)
            node[near start,right] {*}
            node[near start,left] {$reqs$}
            -| (Block.south);
        \draw[composition] (KPI.north) -- ++(0,0.6)
            node[near start,right] {*}
            -| node[near start, above] {$kpis$}
                (Block.south) node[near end,right] {1};
        \draw[composition] (Block.north) -- (Model.south)
            node[near start,left] {*}
            node[near start,right] {$blocks$}
            node[near end,left] {1};
    \end{tikzpicture}
    \caption{\label{fig:metamodel}Metamodel of our language in UML notation}
\end{figure}

The metamodel we use is shown in Figure~\ref{fig:metamodel} and consists of the following modeling elements:

A \textbf{model} is the main modeling unit and contains \textit{blocks}. The whole of Figure~\ref{fig:runingExample} depicts a model, augmented with additional analysis results.

\textbf{Blocks} are the basic hierarchical structuring element of the model. In Figure~\ref{fig:runingExample}, every box with a blue titlebar represents a block, like \texttt{AutonomousDriving} or \texttt{Vehicle}. Each block has a name and contains a list of \textit{properties}, \textit{requirements}, and \textit{kpis}. Blocks can be used to represent arbitrary concepts from the modeled domain, such as components, stakeholders, or services. Given a model~$m$, we say that $blocks(m)$ denotes the set of all blocks within $m$, whereas $children(b) \subset blocks(m)$ is the set of all direct children of a block $b \in blocks(m)$.

Blocks can also be related to each other in an \emph{interface-implementation} association. If block $b$ is an implementation of block $a$, we consider block $b$ as a possible solution for $a$. In this case, block $a$  takes on the role of an interface and block $b$ the role of an implementation. Note that a single block can also simultaneously act as an interface and an implementation by creating multiple such associations. In Figure~\ref{fig:runingExample}, \texttt{Fuse} is an interface implemented by the two blocks \texttt{BlFuse} and \texttt{EFuse}.

We define the set of direct implementations $impls$ of an interface $a$ as:
\[
impls(a) = \{b \in blocks(m) \mid b \text{ implements } a\}
\]

Conversely, we define the set $interfaces$ of all blocks within a model $m$ that take on the role of an interface as:
\[
interfaces(m) = \{ a \in blocks(m) \mid impls(a) \neq \emptyset \}
\]

And we further define the set of \emph{all} implementations $allimpls$ of an interface $a$ as the transitive closure over $impls$:
\[
allimpls(a) = impls(a) \cup\!\!\!\!\!\!\!\!\!\bigcup\limits_{b\,\in\,impls(a)}\!\!\!\!\!\!\!\!allimpls(b)
\]

Interfaces and their respective implementations define points of variability in the model that can be used to capture the structural diversity and inherent uncertainty of roadmapping. \label{change-multiple-interfaces}Each interface forms what we call a \emph{solution space}, as will be explained on page~\pageref{solution-space}, and allows the selection of one of its implementations to act as the default solution. Often, a single block can be part of multiple different solution spaces across a model. To facilitate this we allow each block to implement multiple interfaces. Restricting this to a single interface would require the implementation block to be duplicated instead, which would in turn cause unwanted redundancy. In case of the diamond problem, i.e. if a block $a$ transitively implements the block $d$ through two different paths $b$ and $c$, we resolve conflicting overwrites between $b$ and $c$ by applying them in lexical order. This order is also discernible in the provided user interface within each block, where interfaces are listed top to bottom.

\label{change-nested-blocks-inherit}The \emph{children} and \emph{impls} relations are orthogonal to each other, which brings up the question whether implementations should inherit child blocks from their interfaces. Inheriting child blocks would allow interfaces to propagate a common substructure downwards into all their implementations. This in turn could be combined with an overwriting mechanism in order to allow interfaces to specify required sub-components which should be provided by all implementations. We have experimented with such semantics, and have found the added expressiveness to not be worth the significant increase in complexity. Instead, we have decided against inheriting child blocks. This decision in turn allows a convenient modeling pattern: If the implementations to an interface are only used locally, and not across the whole model, they can be grouped as children of its interface. This reduces the distance between related modeling elements and can improve readability.

\textbf{Properties} contain a name and a formula that describes the value of the property. They are displayed in Figure~\ref{fig:runingExample} as individual entries below a block title. The embedded textual expression language of formulas supports various operations and references to other properties, and will be described in more detail in Section~\ref{sec:exprlng}. Furthermore, expressions can also reference a global time parameter in order to define time-dependant values, as will be described in Section~\ref{sec:time-dependence}.

We define the set of all explicit properties of a block~$b$ as $props(b)$.

In case of an interface-implementation association between two blocks, the implementation implicitly inherits all properties of the interface. This is, for example, the case for the property \texttt{BatteryVoltage} of the block \texttt{BlFuse}, which is inherited from \texttt{Fuse}. If a block defines a property with the same name as an inherited property, the local definition overrides the inherited one. We define the set of all properties $allprops$ (including the inherited ones) of a block $b$ in a model $m$ as:
\[
allprops(b) = props(b) \cup \{allprops(i) \mid b \in allimpls(i),\; \forall i \in interfaces(m)\}
\]

\textbf{Requirements} contain an expression that describes a boolean condition. They are displayed in Figure~\ref{fig:runingExample} as individual rows within a block, and they use the same expression language as properties. Therefore, requirements can reference properties and other aspects of the model, and can perform calculations with their values. The purpose of a requirement is to specify the conditions under which the containing block becomes available. Similar to properties, requirements are also inherited along an interface-implementation association. This is, for example, the case for the requirement \irisinline{MaxLoadCurrent >= Vehicle.TotalCurrent} of the block \texttt{BlFuse}, which is inherited from \texttt{Fuse}. 

The inheritance of a requirement can not be prevented, and an inherited requirement can not be further adjusted. Therefore, inherited requirements can not be relaxed by making the condition less strict, or by removing it completely. The inheriting block can, however, add additional explicit requirements and thereby make the overall requirements for this block stricter.

We define the set of all explicit requirements within a block $b$ as $reqs(b)$, and the set of all requirements $allreqs$ in a model $m$ as:
\[
allreqs(b) = reqs(b) \cup \{allreqs(i) \mid b \in allimpls(i),\; \forall i \in interfaces(m)\}
\]

\textbf{KPIs} (Key Performance Indicators) contain an expression that represents a metric. The running example in Figure~\ref{fig:runingExample} contains such a KPI as the first row \irisinline{num(Watchdog)} in block \texttt{Fuse}. The purpose of a KPI is to define a metric for the quality or fitness of competing solution alternatives. While such a metric might also be useful for roadmap engineers as a form of documentation, or as a data source for strategic decisions, their main purpose is to allow our analysis to perform an automated selection of the best solution alternative. Since expressions and therefore KPIs can reference properties and the global time parameter, such an automated selection can depend not only on the values of properties of the containing and other blocks, but also on different points in time within the roadmap. Section~\ref{sec:time-dependent-modeling-concepts} will describe this aspect in more detail.

Unlike properties and requirements, KPIs are not inherited. This ensures that each interface can define its own relevant set of KPIs, without being affected by selection criteria used in other parts of the model. \label{change-kpi-example}An example for this would be a block, like \texttt{BlFuse}, that is part of multiple different solution spaces within a model. One solution space could represent a required fuse for the engine controller, whereas a different part of the model could require a separate fuse for the headlights. Both solution spaces might consider similar solution alternatives, yet have different selection criteria based on their individual needs. In such a case, inheriting the KPIs would add both sets of selection criteria to the block \text{BlFuse}, even though those criteria are specific to the context of each solution space.

\bigskip

All properties, requirements, and KPIs within a model implicitly define a global system of equations that can be solved using, for example, an SMT Solver (Satisfiability modulo theories, cf.~\cite{10.1145/1995376.1995394}), or our own solving approach based on symbolic transformations, as will be described in Section~\ref{sec-solver}. This system of equations depends on the configuration of blocks active for the currently selected point in time, which in turn depends on the values of properties, requirements and KPIs. In order to resolve these interconnected dependencies, our analysis consists in constructing and solving the global system of equations.

We say that for a given property or requirement $x$, $solve(x)$ denotes the resulting value range of $x$ after solving the equations. In Figure~\ref{fig:runingExample}, the solver results of properties and KPIs are displayed to the right of each formula, whereas the results of requirements are displayed as background colors. Section~\ref{sec-visu} contains a description of different forms of visualization in the model, and the semantics of the color schema.

The combination of interface-implementation associations and requirements allows us to define the notion of \emph{availability} of a block.
We say that a block (or better: the technological concept represented by a block) is available if all of the following criteria are true:

\begin{itemize}
    \item If the block has any implementations, at least one of the implementations must be available.
    \item If the block contains any children, all children must be available.
    \item If the block contains any requirements, all requirements must evaluate to \texttt{true}.
\end{itemize}

The availability $avail$ of a block $a$ is therefore defined as:
\[
avail(a) = \left(impls(a) = \emptyset \vee \!\!\!\!\!\!\!\!\!\!\bigvee\limits_{i\,\in\,impls(a)}\!\!\!\!\!\!\!\!\!\!avail(i)\right) \wedge\!\!\!\!\!\!\!\bigwedge\limits_{c\,\in\,children(a)}\!\!\!\!\!\!\!\!\!\!\!\!\!avail(c) \wedge\!\!\!\!\!\!\!\!\!\!\bigwedge\limits_{r\,\in\,allreqs(a)}\!\!\!\!\!\!\!\!\!\!\!solve(r)
\]

Note that the automated selection mechanism based on KPIs prefers available implementations. If at least one implementation is available, the mechanism will never select an unavailable one. Therefore, in order to determine the availability of a block, it is not necessary to check the availability of the automatically selected solution alternative. Instead we can check that at least one implementation is available.

\label{solution-space}For a given block $b$ representing an interface, we call the set of all available implementations $\{ a \in allimpls(b) \mid avail(a)\}$ the \emph{solution space} of the interface. \label{change-blfuse-available}In our running example, the solution space of the block \texttt{Fuse} contains only the implementation \texttt{BlFuse}, since \texttt{EFuse} is not available due to an unsatisfied requirement.

For a given model $m$, we call a mapping that selects an available implementation from the solution space of each interface $i \in interfaces(m)$ a \emph{configuration} of $m$. Since interfaces act as points of structural variability in our modeling language, such a configuration represents a valid way to \enquote{build} or \enquote{instantiate} the modeled system with actual implementations. The set of all possible configurations defines a \emph{model-wide solution space}. This model-wide solution space is important for roadmap engineers, as it contains all valid configurations of blocks within the model that satisfy the specified requirements.

In order to ease working with such solution spaces, our implementation offers an automated selection mechanism based on the defined KPIs: For each interface in a model, all \textit{available} implementations are ranked according to the sum of all KPIs defined in the interface, and the best alternative is selected as the implementation. \label{why-unavailable-solutions}If none of the implementations are available, however, it can still be important for roadmap engineers to analyze the ranking between those implementations. The unavailability might be caused by erroneous requirements, or the implementations might each fail to satisfy different requirements. These situations, where none of the existing solutions are fully satisfying, provide important insights into the tradeoffs between different alternatives, and can help in identifying future needs. Therefore, our automated selection mechanism also proceeds in cases where all implementations are unavailable. If no KPIs are defined for an interface, then no automated selection is performed for that interface, and the solution space is left open.

In our example, the user specified the KPI \irisinline{num(Watchdog)}, which evaluates to $0$ or $1$ depending on the value of \verb|Watchdog|. The function call to \irisinline{num} ensures that boolean values are converted to numbers, which is required by our automated selection mechanism in order to compare different KPI values. If both alternative implementations were available, the system would prefer the one with a watchdog. However, since \texttt{EFuse} is not available in the example, \texttt{BlFuse} is selected.

The automated selection mechanism based on KPIs is all the more important when introducing time-dependence into the model, as the user would otherwise have to make a manual selection of implementations at each point in time. In this case, the model-wide solution space also becomes time-dependent. If the user specifies a point in time (via the time slider that we provide in our user interface), and manually selects solution alternatives for all interfaces that were not automatically selected via KPIs, then the resulting visible model represents a single configuration. If, instead, one or more solution spaces are left undecided, then the visible model represents a set of configurations. Furthermore, if the user wants to inspect the set(s) of configurations defined represented by the model across all points in time, then this is possible using several visualizations in the user interface, like plots showing values over time, or a chart overview that shows the time-dependent availabilities of all blocks within a model (c.f. Section~\ref{sec-visu}).

\subsection{Expression Language}\label{sec:exprlng}

We use an embedded textual expression language for specifying properties, requirements, and KPIs.
The language is expressive enough to represent complex formulas, while still being simple enough to be written and understood by domain experts. Since Microsoft's office suite, especially Excel, is widely used in industry as reported by our industrial partners, the syntax and functionality of our language is similar to Excel formulas. Our aim is to enable  domain experts to easily apply their existing skills to our roadmap modeling and analysis approach.

This intended simplicity led us to avoid existing languages like OCL~\cite{OCL2014} due to their complexity and the effort required to implement the additional usability features that will be explained in Section~\ref{sec-visu}. The language provides a set of literals, including literals to specify SI-units, identifiers, arithmetic and relational operators, conditional expressions, aggregation expressions, and function calls. \ref{app:syntax} describes the individual language elements in more detail. Furthermore, Section~\ref{sec:time-dependent-properties} will extend the language to add more time-related constructs. 

The combination of our structural modeling languages and the embedded expression language allows us to create a model representing the interconnected composition of technologies relevant for technology roadmapping. Such a model, however, does not yet support any variability between different points in time.

\subsection{Time Dependence}\label{sec:time-dependence}

In our context, the term \emph{time} refers to an arbitrary point in time in the future, when the model is supposed to be relevant.
A concrete point in time is specified by a date.
Since we use modeling primarily for roadmapping of new technologies, looking at a period of 10~-- 30 years in the future, we use a granularity of years and months for our dates.

This definition of the concept of time differs from the clock-based \emph{realtime} usually found in formal models, like in Time Petri Nets~\cite{CR05} or Timed Automata~\cite{Alur&Dill1994}, or in realtime extensions of the UML~\cite{OMG2009}.
As both definitions of time can play a role in our models, e.g., when specifying requirements for the reaction time of a system in the year 2030, we use the terms \emph{roadmap time} and \emph{runtime} to distinguish between the two implied meanings of the term. 

We further define two forms of time-dependence:

\begin{enumerate}
    \item \textit{Local Time Dependence}: the value of a property changes over time. This can be useful, for example, to model the predicted increase of power consumption of car components over time, or the decreasing legal thresholds on CO$_2$ emissions over time.
    \item \textit{Structural Time Dependence}: structural aspects of the model change over time. At different points in time, the structural composition of the model can be different, containing different sets of blocks. This can be useful, for example, to model emerging technologies, like the availability of reversible smart fuses.
\end{enumerate}

In the following we describe how our modeling language has been extended to support both forms of time-dependence.

\subsubsection{Local Time Dependence\label{sec:time-dependent-properties}}

The values of all properties in our model are defined by their respective formulas.
In order to express time dependence within properties, we add an implicit parameter $T$ to each property, representing an arbitrary but fixed point in time.
This parameter $T$ can be referenced within the formula to make the resulting value of the property time dependent, by using the literal \irisinline{T}. This literal \irisinline{T} is, for example, used in Figure~\ref{fig:runingExample} in the property \texttt{TFLOPS} of the block \texttt{DetectionSoftware}.

When solving the whole model, we instantiate all properties within the model with a global user-selected value for $T$.
In our implementation, this value is specified by the user setting a time-slider in the UI, as will be shown in Section~\ref{sec:time-dependent-modeling-concepts}.

We further extend our expression language with additional capabilities to work with time.
For this, we make \emph{time} a first-class citizen of our expression language by defining two new types of values, in addition to boolean values and numbers:

\begin{itemize}
    \item A \emph{date} specifies a single point in time.
    \item A \emph{duration} specifies the length of a time interval.
\end{itemize}

Both value types can be created by respective literals in our embedded expression language:

\begin{itemize}
    \item \emph{Date literals} specify a month name and year in the form \irisinline{MonthYear}, e.g., \irisinline{Feb2035}.
    \item \emph{Duration expressions} specify a number of months in the forms \irisinline{months(n)} and \irisinline{years(n)}. For example, \irisinline{months(24)} and \irisinline{years(2)} both specify a duration of 24 months.
\end{itemize}

We distinguish between these two types of time values in order to provide different textual representations for the user (e.g. \textit{2021 years 7 months}  for a duration vs. \textit{Aug2021} for the corresponding date), and in order to define unambiguous arithmetic operations.
For example, multiplying a date by the number 2 is ambiguous, as it depends on the neutral element of the applied calendar system (i.e., year 0).
Multiplying a \emph{duration} by the number 2, however, is well-defined, as it simply doubles the number of months specified by the duration. \ref{app:date-ops} lists all arithmetic and relational operators that we support on dates and durations. \label{old:date-ops}

In addition to these changes to the expression language, we also extend the meaning of identifiers.
Each identifier, referencing either a property or a block availability in the model, can now be suffixed by an additional time argument, e.g., \irisinline{TFLOPS(Jan2030)}.
If no argument is specified, the implicit argument \irisinline{(T)} is used.
This makes it possible to query the value of properties or the availability of blocks at a time different from the current $T$.

\subsubsection{Structural Time Dependence}

Time dependent properties do not affect the structure of the model, i.e., the defined set of blocks and their associations.
In the context of roadmapping, however, we want to specify different configurations of blocks at different points in time, using a single model.
This allows us to express, for example, that we predict new technologies to become available in the future, or that existing technologies become unavailable due to changes in legislation.

In order to allow the structure to change depending on the selected point in time, we make use of the concept of availability.
The availability of a block is based on the satisfiability of its requirements.
By adding the same parameter~$T$ to requirements that we also added to properties, we can in consequence vary the availability of a block over time.
In our running example in Figure~\ref{fig:runingExample}, this is used in the block \texttt{EFuse}, where the requirement at the bottom states that the whole block is unavailable before January 2022 plus an additional time frame of 12 to 36 months (see Subsection~\ref{subsec:interval-arithmetic} for more details).
As another example, by adding a requirement \irisinline{A(T - years(2))} to a block \texttt{B}, we cause block \texttt{B} to become available two years after block \texttt{A} becomes available. With this requirement, the availability of block \texttt{B} at any point in time \irisinline{T}, e.g. at \irisinline{T = Jan2022}, is equal to the availability of block \texttt{A} two years prior, i.e. at \irisinline{Jan2022 - years(2) = Jan2020}.

The combination of time-dependent properties, requirements, KPIs, and availabilities leads not only to a time-dependent model, but also to a time-dependent solution space. When instantiating a model with two different values of $T$, all properties, requirements and KPIs within the model can evaluate to different values. This in turn can lead to different outcomes for all block-availabilities, and therefore different solution spaces for each interface. The resulting two model slices can therefore differ significantly from each other, both in terms of local values, and global structure.

\subsection{Dealing with uncertainty}\label{sec:uncertainty}

\label{subsec:interval-arithmetic}When working with time-dependent data and models from real-world applications, many values are not known exactly in advance, yet constrained to lie in a certain range.
Thus the modeling language provides interval-semantics as a way
of expressing those uncertainties.

For instance, in our running example, the development of the EFuse is planned to start in January 2022. Yet, because of some unknown factors, the effective time-to-market is not known beforehand and estimated to require between 12 and 36 months. In our modeling language, we can express this with the expression \mbox{\irisinline{T >= Jan2022 + [months(12)..months(36)]}}.

With \irisinline{[months(12)..months(36)]} we define an interval with a \textit{lower} and an \textit{upper} bound, represented by \texttt{[\textit{lower}..\textit{upper}]}. This interval defines the mathematical value range between the given bounds as a closed interval in \(\mathbb{R}\): \([\text{lower},\;\text{upper}]\).
Because we use those intervals to represent uncertainties, the
 expression \mbox{\texttt{x = [\textit{lower}..\textit{upper}]}} is not to be mistaken as \texttt{x} \emph{being} the interval \([\text{lower},\;\text{upper}]\).
For intervals are used to represent value ranges, the expression \texttt{x = [\textit{lower}..\textit{upper}]} is to be interpreted as a constraint for \texttt{x} to lie in the given range.

In order to adequately support intervals, all basic arithmetic operations are implemented as follows.
All of them are inclusion isotonic~\cite{moore1966interval},
meaning that for the intervals \(A_1\), \(A_2\), \(B_1\) and \(B_2\) all interval operations \(\circ\) satisfy the requirement:
\begin{equation*}
    B_1 \subseteq A_1,~ B_2 \subseteq A_2 \implies B_1 \circ B_2 \subseteq A_1 \circ A_2
\end{equation*}
Albeit this is easy to ensure for some operations, the division is more difficult as the divisor may contain \(0\). Consult \ref{app:interval-ops} for more details.\par
{%
\let\varstyle\texttt\def\a{\varstyle{a}}\def\b{\varstyle{b}}\def\c{\varstyle{c}}\def\d{\varstyle{d}}%
\def\T#1#2{\texttt{[\ensuremath{#1}..\ensuremath{#2}]}}%
\def\RT#1{\texttt{[\ensuremath{#1}..\(\infty\))}}%
\def\LT#1{\texttt{(\(-\infty\)..\ensuremath{#1}]}}%
\def\InfT{\texttt{(\(-\infty\)..\(\infty\))}}%

Alongside the arithmetic operations, intersections are implemented as well. If \(\exists x \in \mathbb{R}: x \in \T{\a}{\b} \land x \in \T{\c}{\d}\), the resulting interval is defined as: \begin{equation*}
    \T{\a}{\b} \cap \T{\c}{\d} = \T{\max\{\a,\c\}}{\min\{\b,\d\}}
\end{equation*}
Otherwise (that is, if the intervals are disjoint) the resulting interval is empty. Therefore \(\T{2}{4} \cap \T{3}{7}\) evaluates to \(\T{\max\{2,3\}}{\min\{4,7\}} = \T{3}{4}\) while \(\T{1}{2} \cap \T{3}{4}\) evaluates to the empty interval.
Since intervals denote a range of values, the empty interval will be tainted to flag an invalid value: A property that cannot evaluate to any value is invalid and therefore all of its usages are invalid too.
To propagate the erroneous value, any operation performed with an empty interval will result in an empty interval (and therefore a tainted value) again.

Constraints like \mbox{\irisinline{[a..b] >= c}} are implemented as follows:
\begin{alignat*}{2}
    (\T{\a}{\b}     &\geq     \c) &&= (\T{\a}{\b} \cap \RT{\c})\\
    (\T{\a}{\b}     &\leq     \c) &&= (\T{\a}{\b} \cap \LT{\c})
\end{alignat*}%
Furthermore, the utility functions (\(\sin\), \(\cos\), \ldots) have been extended to support intervals. With \texttt{PI} being a constant for the mathematical constant \(\pi\), expressions like \irisinline{min(sin([-PI..PI]/3), 5, [-1..3])} evaluate to the value range \T{-1}{\sin(\pi/3)} and non-monotonic functions (like \(\sin\) and \(\cos\)) are not just evaluated based on corner cases, since that could possibly miss local minima\nobreak /\allowbreak maxima. Instead we perform an analytically correct computation equivalent to taking the union of all possible outputs when using the whole input domain.

\bigskip

Having intervals only representing the value range of a property,
comparisons like \irisinline{x == 3} are not just bound to boolean logic.
If the closest value range for \texttt{x} is the interval \T{0}{4},
\irisinline{x == 3} should yield neither \textit{true} nor \textit{false}.
Therefore the boolean logic is extended by the uncertainty value \textit{maybe}, which can be viewed as the Interval \irisinline{[false..true]}. All three important operations (not, and, or) are defined with the following table, similar to Kleene's strong logic of indeterminancy:
\begin{table}[H]
    \centering\ttfamily\begin{tabular}{cc@{\hspace{2.25em}}ccc}
        \toprule
            A     & B     & !A    & A \& B & A | B \\
        \cmidrule(r{2em}){1-2}\cmidrule{3-5}
            false & maybe & true  & false  & maybe \\
            maybe & false & maybe & false  & maybe \\
            true  & maybe & false & maybe  & true \\
            maybe & true  &       & maybe  & true \\
            maybe & maybe &       & maybe  & maybe \\
        \bottomrule
    \end{tabular}
\end{table}
As a result, the comparison \irisinline{[1..3] = [0..2]} evaluates to \textit{maybe}, but \irisinline{[-2m..1m] <= [5m..6m]} evaluates to \textit{true} because even the maximum possible value on the left hand side (being 1 meter) is lower than the minimum possible value on the right hand side (being 5 meters). More precisely, the equal-comparison of two intervals is \textit{true} if and only if both intervals contain exactly one value (e.g. \irisinline{[2m..2m]}) and this value is the same for both intervals (if the value differs, the comparison yields \textit{false}). If at least one of the intervals contains more than one number (e.g. \irisinline{[0m..12m]}), the comparison yields \textit{maybe} if both intervals overlap and \textit{false} if they are disjoint.
}

\section{Generating and solving the constraint system}\label{sec-solver}

A model as defined through the presented meta-model includes several distinct semantic aspects, which need to be taken into account when deciding on an evaluation technique:
\begin{itemize}
    \item Inheritance relations implicitly define additional model elements.
    \item Hierarchical nesting between blocks affects availabilities and identifier scoping.
    \item Requirements affect availabilities.
    \item Property values depend on the active solution alternatives.
    \item Automated choice between solution alternatives is determined by comparing different KPIs and availabilities.
\end{itemize}

In order to avoid dealing with all these aspects at once, we first reduce the complexity of the model in the following phases by transforming it into intermediate forms of decreasing complexity:

\begin{enumerate}
    \item The model is expanded by explicitly creating properties and requirements that are defined implicitly by inheritance relations. After this phase, all relevant model elements exist explicitly.
    \item Identifiers in expressions are resolved based on hierarchical relations. After this phase, the hierarchy within a model can be ignored apart from its effects on availabilities.
    \item The model is lowered into a flat constraint system. This phase abstracts away the remaining aspects, producing a form that can be processed without having to take into account availabilities and solution spaces.
\end{enumerate}

The constraint system is then solved using repeated symbolic transformations, which result in a simplified constraint system containing value bounds for the various model elements. Afterwards, the resulting value bounds are displayed in the user interface, as will be shown in Section~\ref{sec-visu}.
The following sections explain in more detail the three transformation phases, as well as the actual solving process of the constraint system.

\subsection{Expanding inheritance}

This phase works by walking along all inheritance relations in the model and carrying over properties and requirements. Topological sorting is used to correctly handle transitive inheritance. In case of properties, explicitly defined properties overwrite inherited ones with the same name. If a property name is inherited through multiple inheritance relations, then the deterministic lexical order of relations is used to resolve conflicts. Requirements are inherited unconditionally. KPIs, however, are not inherited. Since KPIs are used to assist an automated strategy selecting between different solution alternatives, such KPIs should stay local to the context where a solution alternative is applied, and not where it is defined.

In our running example, the inheritance relation between the blocks \irisinline{Fuse} and \irisinline{BlFuse} results in the creation of the inherited requirement \irisinline{MaxLoadCurrent >= Vehicle.TotalCurrent}, and the property \irisinline{BatteryVoltage}. The other two properties \irisinline{MaxLoadCurrent} and \irisinline{Watchdog} are overwritten in the block \irisinline{BlFuse}. The KPI \irisinline{num(Watchdog)} is not inherited.

The expansion of inheritance relations creates independent copies of model elements, which is important, because inherited properties often result in a different value than their ancestors due to referenced properties being overwritten.

\subsection{Resolving identifiers}

After the expansion of inheritance relations, all identifiers used in expressions throughout the model are resolved to their respective model elements. This resolving is performed based on static scoping rules determined by the block hierarchy: names are first looked up locally within the respective block, and, if not found, recursively within its hierarchical ancestors.

Almost all expressions are resolved within the scope of their containing block. The only exception is for KPI expressions, which need to be resolved separately within the scope of each solution alternative. This ensures that the expression measures the quality or fitness of the solution alternative, and not of the interface itself.

Special care needs to be taken for aggregations like \texttt{SUM(\textit{expr})}. These expressions compute a function over all direct descendant blocks of a common parent block. In order to \enquote{resolve} the hierarchical relationships, we replace aggregations with their expanded form. Therefore the implementation of each aggregation does not perform the computation itself, but instead performs the transformation that lowers the aggregation into a semantically equivalent expanded expression.

In our running example, the aggregation \irisinline{SUM(Current)} in the property \irisinline{Vehicle.TotalCurrent} is replaced with the expression \texttt{Headlights.Current + ProcessingUnits.Current}, which in turn gets resolved to the actual properties.

After this phase, the hierarchical relationships within a model can be ignored, apart from its effects on availabilities.

\subsection{Generating the constraint system}

Finally, the expanded and resolved model is converted into a flat constraint system. This constraint system consists of a set of true statements, i.e. expressions in our expression language that must always evaluate to \irisinline{true}.

Given the capabilities of our expression language, we can already use it to express such statements over property values, like for example \irisinline{A.P(T) = 10}, meaning that the value of property \texttt{P} of block \texttt{A} at time \texttt{T} is \(10\). However, this is not yet sufficient to generate a constraint system that captures the full semantics of our modeling language. For example, we cannot express a constraint stating that a block becomes unavailable if one of its requirements is not satisfied because we have no means to reference model aspects like availabilities. To address this, we extend our expression language to allow the following types of references:

\begin{itemize}
    \item References of the form \irisinline{A.?requirementN(T)}, which represent the value of the \texttt{N}-th requirement of block \verb|A| at time \verb|T|.
    \item References of the form \irisinline{A.?kpiN(B, T)}, which represent the value of the \texttt{N}-th KPI of block \verb|A|, when evaluated in the context of the derived block \verb|B| at time \verb|T|.
    \item References of the form \irisinline{A.?availability(T)}, which represent the availability of block \verb|A| at time \verb|T|.
    \item References of the form \irisinline{A.?replacement(T)}, which represent the index of the active solution alternative of block \verb|A| at time \verb|T|, or the value \verb|-1| if block \verb|A| has no solution alternatives.
\end{itemize}

The prefix \enquote{\texttt{?}} ensures that the new references to not clash syntactically with user defined block or property names.

In order to create the initial constraint system, we iterate over all blocks, properties, requirements, and KPIs in the resolved model and generate one or more constraints for each of these model elements as follows:

For each \textbf{property} $\texttt{P} \in allprops(\texttt{A})$ of block \verb|A| with expression \verb|<expr>| we generate a constraint that ensures that the value of \irisinline{A.P} at time \irisinline{T} is equal to \verb|<expr>|. In the simple case where \verb|A| has no solution alternatives, i.e. $allimpls(\texttt{A}) = \emptyset$, this could be achieved with the constraint \irisinline{A.P(T) = <expr>}. However, if block \verb|A| does have solution alternatives $allimpls(A) = \left\{ \texttt{B\_1}, \ldots, \texttt{B\_N} \right\}$, the value of the property should instead be equal to the value of the corresponding property \verb|B_k.P| in the selected solution alternative. In order to handle both cases uniformly, we generate the following constraint:

\begin{iris}
A.P(T) = (if (A.?replacement(T) = 1) then B_1.P(T)
     ...
     else if (A.?replacement(T) = N) then B_N.P(T)
     else <expr>)
\end{iris}

\noindent
This expression maps the property \verb|A.P| to one of the properties \verb|B_k.P|, with \verb|k| determined by the value of \verb|A.?replacement|.

Note, that if the expression \verb|<expr>| refers to \verb|T|, then it uses the time parameter provided to \verb|A.P| instead of the global time. This makes it possible to refer to the property \verb|A.P| at different points in time by writing, for example, \irisinline{A.P(Jan2030)}.

For each \verb|N|-th \textbf{requirement} $\texttt{R} \in allreqs(\texttt{A})$ of block \verb|A| with expression \verb|<expr>| we generate the constraint \irisinline{A.?requirementN(T) = <expr>}. The selection process based on \verb|A.?replacement|, that we used for properties, is not necessary for requirements, since requirements cannot be overwritten in derived blocks.

For each \verb|N|-th \textbf{KPI} $\texttt{K} \in kpis(\texttt{A})$ of block \verb|A| with expression \verb|<expr>|, we generate a constraint for each solution alternative $\texttt{B\_i} \in allimpls(\texttt{A})$ of block \verb|A|. The constraint has the form \irisinline{A.?kpiN(B_i, T) = <expr>}, where \verb|<expr>| is resolved within the scope of \verb|B_i|. Therefore, the value of \irisinline{A.?kpiN(B_i, T)} is equal to the value of the KPI metric evaluated for the solution alternative \verb|B_i|.

For each \textbf{block} \verb|A| we generate two additional constraints to handle availabilities and the automated selection of solution alternatives based on KPIs.

In order to handle the \textbf{availability} of block \verb|A|, we generate the following constraint:

\begin{iris}
A.?availability(T) =
  (     if (A.?replacement(T) = 1) then B_1.?availability(T)
        ...
   else if (A.?replacement(T) = N) then B_N.?availability(T)
   else true)
  & (A.?requirement1(T) & ... & A.?requirementM(T))
  & (C_1.?availability(T) & ... & C_K.?availability(T))
\end{iris}

Here, $\left\{ \texttt{B\_1}, \ldots, \texttt{B\_N} \right\} = allimpls(\texttt{A})$ denote the solution alternatives of block \verb|A|, $\texttt{M} = \left\vert allreqs(\texttt{A}) \right\vert$ is the count of requirements in block \verb|A|, and $\left\{ \texttt{C\_1}, \ldots, \texttt{C\_K} \right\} = children(\texttt{A})$ are the direct children of block \verb|A|. The constraint states that block \verb|A| is available iff the selected solution alternative is available, and all requirements are satisfied, and all of its direct descendants are available. \label{change-block-available-if-no-selection}If block \texttt{A} has no implementations, then the first part of the constraint collapses to \texttt{true}, and the availability is solely determined by its requirements and descendants.

In order to realize the automated selection of solution alternatives based on \textbf{KPIs}, we add the following constraint for each block \verb|A|:

\begin{iris}
A.?replacement(T) = index_of_max(
  if B_1.?availability(T)
     then (A.?kpi1(B_1, T) + ... + A.?kpiM(B_1, T))
     else -inf,
  ...,
  if B_N.?availability(T)
     then (A.?kpi1(B_N, T)) + ... + A.?kpiM(B_N, T))
     else -inf
)
\end{iris}

Again, $\left\{ \texttt{B\_1}, \ldots, \texttt{B\_N} \right\} = allimpls(\texttt{A})$ denote the solution alternatives of block \verb|A|. The parameter $\texttt{M} = \left\vert kpis(\texttt{A}) \right\vert$ represents the count of KPIs in block \verb|A|. The function \verb|index_of_max| returns the 1-based index of its (first) largest argument. We use this to determine the index of the solution alternative which results in the largest KPI value. The if-conditions make sure that only currently available solution alternatives are considered, unless none are available. In that case, the result is simply the first solution alternative. If block \verb|A| has no solution alternatives, then we add the constraint \irisinline{A.?replacement(T) = -1} instead.

\ref{appendix:constraint-system} shows a listing of the whole constraint system that is generated for our running example in Figure~\ref{fig:runingExample}.

The presented rules to generate constraints allow us to fully encode the automated process to select the best solution alternative based on availability and KPIs in the constraint system. The advantage of such an approach is, that after generating the constraint system, further analysis steps can ignore the semantics of hierarchy, inheritance, availabilities, and KPIs. Instead, the analysis can focus on solving the plain constraint system, which greatly reduces complexity.

\label{change-cyclic-dependencies}Each constraint in the generated constraint system is formulated as an equality (\irisinline{=}), with a reference on the left hand side, and an expression on the right hand side. These equalities cannot be solved by processing them like variable assignments in programming languages, where the result of the right hand side is assigned to the variable on the left. In general there exists no strict order of evaluation, since the references used in multiple constraints can form complex dependency graphs, or even cycles. The following section describes our approach to solve such constraint systems.

\subsection{Solving the constraint system}

Given the target user group of roadmap engineers, we have two objectives in solving the constraint system:

\begin{enumerate}
    \item We want to determine the set of possible values (including ranges due to uncertainties) for all properties, requirements, availabilities, and solution alternative selections within the model.
    \item We want to provide tracing information for all results in order to assist roadmap engineers to identify problems and opportunities within the model.
\end{enumerate}

Determining the set of all possible value ranges could be achieved with existing tools like SMT solvers (Satisfiability modulo theories, cf.~\cite{10.1145/1995376.1995394}). Sprey, Sundermann et al.~\cite{10.1145/3377024.3377036}, for instance, have applied SMT solvers in a similar setting to determine attribute ranges for extended feature trees~\cite{10.1016/j.is.2010.01.001} by performing individual optimization analyses to find the lowerst and highest possible values. Extended feature trees share some similarities with our model language by allowing features in a feature tree to have attributes with assigned values. However, adapting such an SMT-based approach to our modeling language semantics would require support for interval arithmetic in order to solve constraint systems in the presence of uncertain value ranges (cf.~\cite{neumaier_1991}). Furthermore, the findings by Sprey and Sundermann~\cite{Sprey2018ComputingAR} suggest that the performance of performing multiple SMT-based optimization analyses per model version is not sufficient for an interactive workflow.

Our second objective, the need to generate tracing information for the user, is difficult to achieve with existing solutions, as this would usually require performing additional solver analyses to extract, for example, a minimal (un)satisfiable core. Another possible approach could be to extract tracing information from proof objects, which some SMT solvers like Z3~\cite{DBLP:conf/tacas/MouraB08} are able to produce~\cite{DBLP:conf/lpar/MouraB08}. Such proof objects describe the sequence of steps and transformations necessary to produce the solver result. However, the generated proof objects are intended for external verification with theorem provers, and it is unclear whether it would be possible to extract our required tracing information through post processing. This led us to implement and integrate our own solution for solving the generated constraint system based on symbolic transformations and the interval arithmetic as presented in Section~\ref{sec:uncertainty}.

In order to find all value ranges, we simplify the constraint system by repeatedly applying symbol transformations on all constraints until a fix-point is reached. \label{change-nontermination}This approach does not always terminate, e.g. in cases where the solution continuously approaches a fix-point with increasing precision without ever reaching it, or in cases where our transformations generate increasingly large constraints. To accommodate this, our solver operates in rounds and sets a threshold (default: 50) on the number of rounds. If the threshold is reached, the bounds inferred up to that point are not wrong, but form a safe super-set, because the solver starts with infinitely large bounds and continuously narrows them down with each inferred constraint. This means that if not all of the information contained within the initial constraint system can be processed, the resulting bounds might contain values that do not actually solve the constraint system. However, the bounds will never exclude valid results.

The goal of all transformations is to achieve a normal form, where constraints are \enquote{\texttt{=}}-relations with a single reference on the left side, and an interval on the right side. The interval on the right can then be displayed in the user interface.

To achieve this, we employ different types of symbolic transformations commonly found in theorem provers~\cite{DBLP:books/sp/NipkowPW02}, like folding constant expressions, propagating inferred value ranges, and using algebraic properties like associativity and commutativity to create new opportunities for simplification. Examples for different classes of transformations that we employ are given in \ref{sec:symbolic-trans}.

All symbolic transformations preserve the correctness of the constraint system, which means that all value assignments that were valid before a transformation are also valid afterwards. However, in some cases transformations are allowed to introduce a relaxation, meaning that after applying the transformation the resulting constraint system might allow more satisfying assignments than before. This happens, for example, when merging resulting value ranges of interval operations. The resulting constraint system therefore represents a conservative over-approximation of the model. This is also the case if the final constraint system contains constraints that did not reach normal form, and therefore do not further restrict the inferred bounds.

\begin{figure*}[tph]
    \centering
    \includegraphics[width=\textwidth,height=.8\textheight,keepaspectratio]{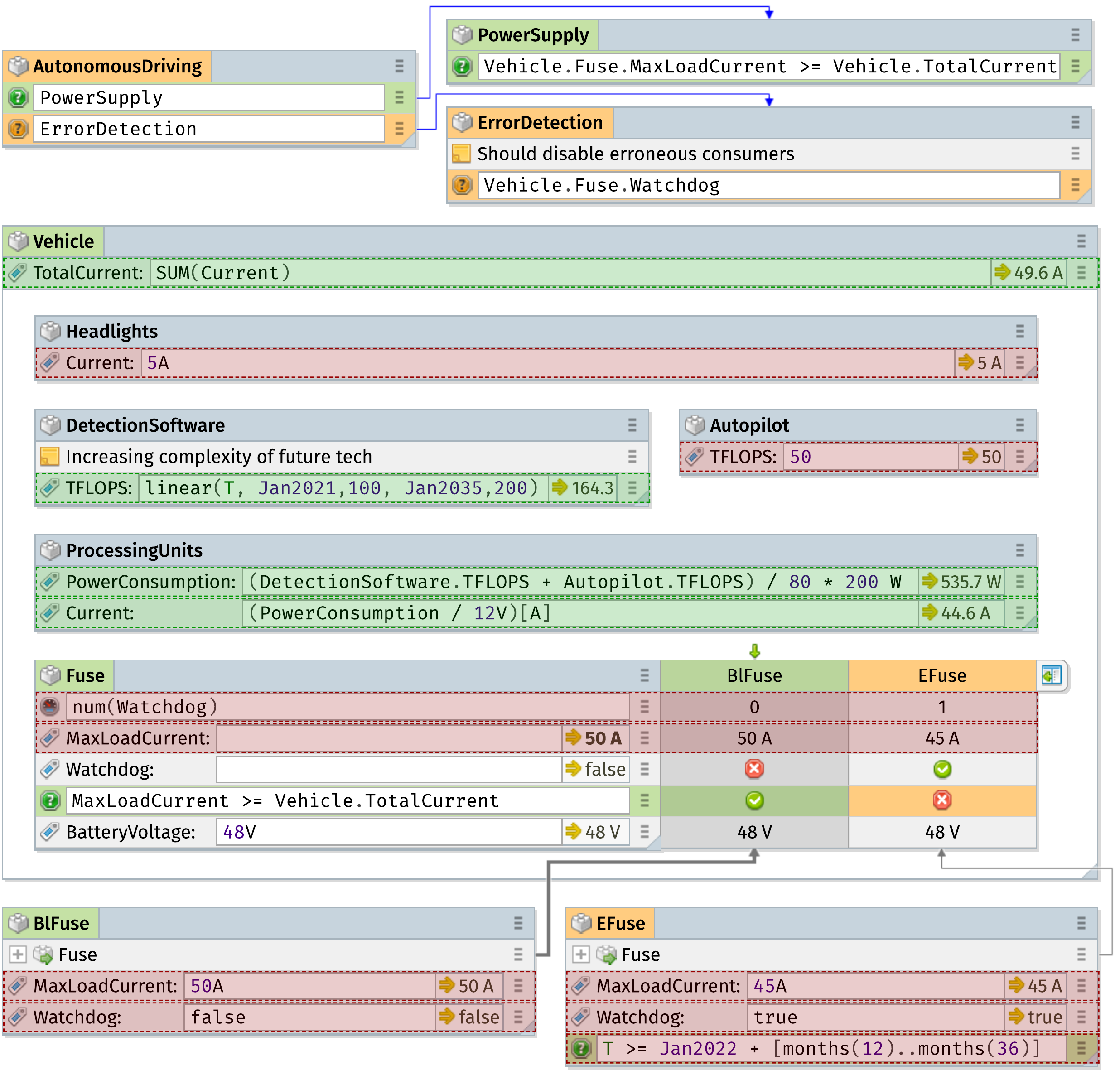}
    \caption{Screenshot of solver traces for the computation of \irisinline{Fuse.MaxLoadCurrent} highlighted in red and green.}
    \label{fig:tracehighlighting}
\end{figure*}

Every application of a transformation rule annotates the resulting expression with the constraints utilized in the transformation. After solving the constraint system, we follow these trace annotations recursively to collect all relevant initial constraints for each resulting value range. This is similar to program slicing~\cite{DBLP:journals/jpl/Tip95} in the area of program language analysis and enables us to visualize the information through the user interface by clicking on a displayed value. This can be seen in Figure~\ref{fig:tracehighlighting}. It enables the user to understand which syntax elements have influenced  the clicked value.

In this visualization, a trace is shown for the value \irisinline{50A} of the property \texttt{Fuse.MaxLoadCurrent}. The tracing information helps users identify all model elements that contribute to an inferred value. When solving the constraint system, each newly created constraint is annotated with the set of all previous constraints and model elements that have contributed to its creation. Every time a constraint $c$ reaches normal form and affects the bounds of the property \texttt{Fuse.MaxLoadCurrent}, the resulting bounds are annotated with the set of transitive dependencies of $c$. This set of transitive dependencies includes all model elements that contribute, in one way or another, to the resulting value \irisinline{50A}. Since the value \irisinline{50A} depends on the automatic selection of \texttt{BlFuse} over \texttt{EFuse}, and this selection in turn depends on the satisfiability of the requirements in \texttt{BlFuse} and \texttt{EFuse}, the traces also contain all model elements necessary for the computation of \texttt{Vehicle.TotalCurrent}.

The traces are highlighted in the user interface as green and red boxes. Elements highlighted in green changed their value from \irisinline{T - 1} to \irisinline{T}, whereas elements highlighted in red remained constant. This can be used by roadmap engineers to identify critical property changes that caused an unwanted value in the model, by setting the current time \irisinline{T} to the first point in time where the unwanted value occurred, and then looking only at those model elements highlighted in green.

In terms of performance and scalability, our approach provides quick results~(\SI{<1}{\second}) for all examples used during our tests and the evaluation reported in Section~\ref{sec:evaluation}, and supports an interactive workflow during typical modeling activities. However, the solver approach does not scale well to larger models without adaptation, where long dependency cycles between expressions would cause large intermediate expressions. Nonetheless, we have found the performance of our approach to be sufficient for our purposes while it additionally provides the added value of the traceability.

Finally, in terms of functional correctness, we employ a combination of manual and automated system and integration tests to verify the implementation of our solver. To this end, our current test suite consists of about 950 automated tests of varying granularity that reach a statement/branch coverage of \SI{92}{\percent}/\SI{91}{\percent} within the ternary logic and interval arithmetic, \SI{78}{\percent}/\SI{70}{\percent} within our symbolic transformations, and \SI{75}{\percent}/\SI{70}{\percent} within the surrounding solver. On top of that, all code is fully typed with TypeScript operating in strict-mode, which ensures a level of type-safety that would otherwise be hard to achieve in JavaScript.

\section{Visualization of time dependent modeling concepts}\label{sec-visu}

Technical models contained in a model-based technology roadmap can be calculated and evaluated by a computer, but roadmapping and the technology selection based on it, as well as the resulting projection to a company's strategy, is done by human (domain) experts. Hence, various challenges arise to make a time dependent model practical and readable for human users.

Most of our visualization techniques focus on the explainability of time-dependent aspects of the model. The variability of the model along the time dimension creates several challenges for an intuitive visual representation. In the following, we describe each challenge and our proposed solution. All screenshots shown in this paper are taken from our prototype \TextIris\ which is available online.\footnote{\label{iris-url}\url{https://genial.uni-ulm.de/jss2021/}}

\subsection{Target User Group}

Through our industrial partners we identified some typical users of a model-based roadmapping tool: In addition to systems engineers, technology scouts and technical experts can also use such a tool. All these various users have in common a deep knowledge of their own domain but might not be modeling experts. Therefore, we decided to choose a graphical syntax for our modeling language in terms of taking up existing well-known graphical languages, such as SysML~\cite{OMGSysML15} and lowering the entry level for non-modeling experts. Furthermore, we implemented the expression syntax as introduced in Section~\ref{sec:exprlng}, inspired by spreadsheet tools.

In the following, we refer to all different types of users mentioned before and those interacting with the technology roadmap as \textit{roadmap engineer}.

\begin{figure}[htb]
    \includegraphics[width=\linewidth]{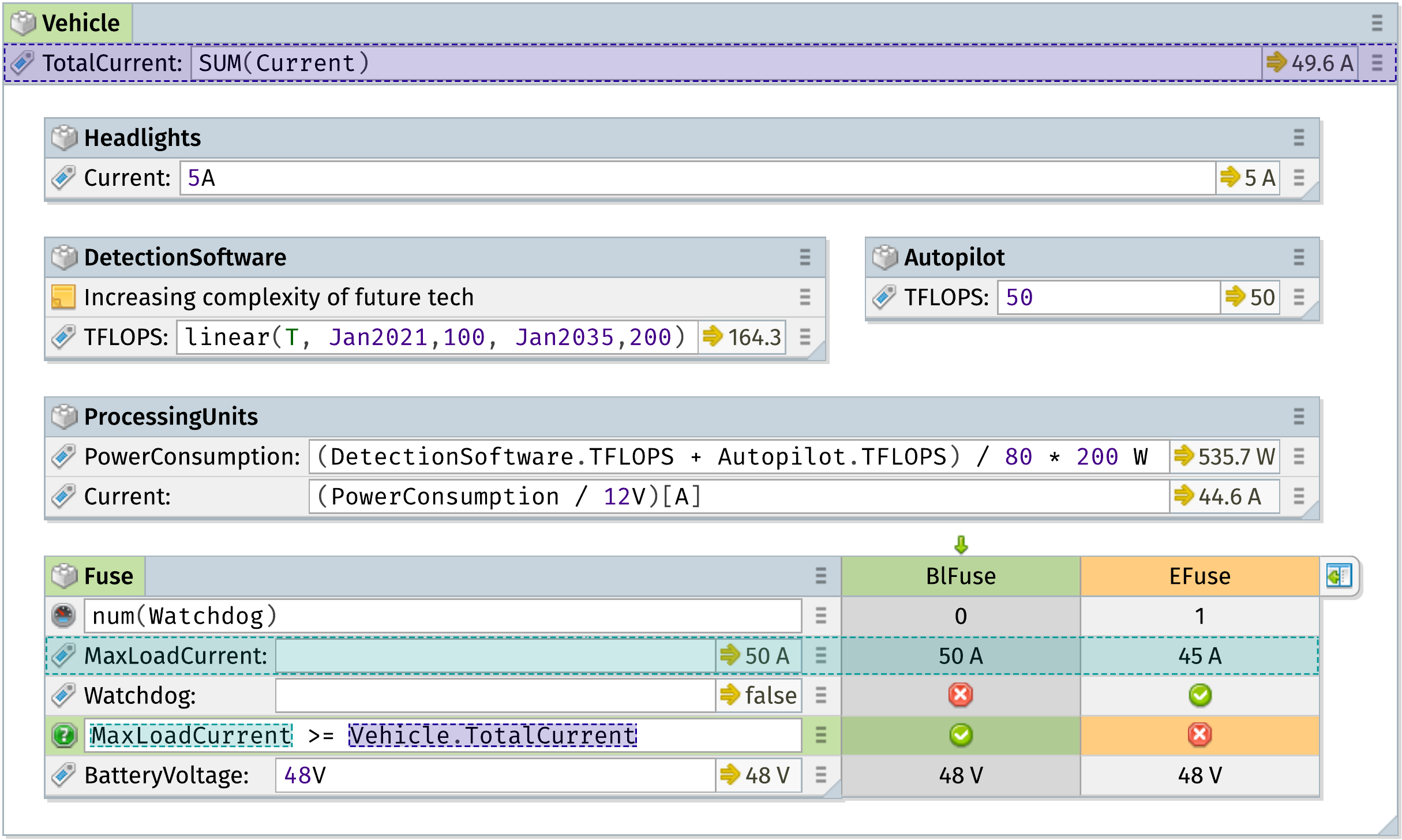}
    \caption{Reference highlighting supports tracing. Currently highlighted are explicit references of requirement \irisinline{MaxLoadCurrent >= Vehicle.TotalCurrent}.}
    \label{fig:referenceHighlighting}
\end{figure}

\subsection{Reference Highlighting}

All properties and requirements of a block can either be a simple value such as \SI{5}{\A} (see block \irisinline{Headlights} in Figure~\ref{fig:runingExample}), or a formula containing references to other properties. To support easy tracing of references, we added color-based highlighting of referenced properties when a roadmap engineer clicks on a formula, as shown for the references which are part of the requirement formula \irisinline{MaxLoadCurrent >= Vehicle.TotalCurrent} in Figure~\ref{fig:referenceHighlighting}.

\subsection{Manual Time Selection}\label{sec:time-dependent-modeling-concepts}
\begin{figure}[htb]
        \centering
        \includegraphics[width=0.85\linewidth]{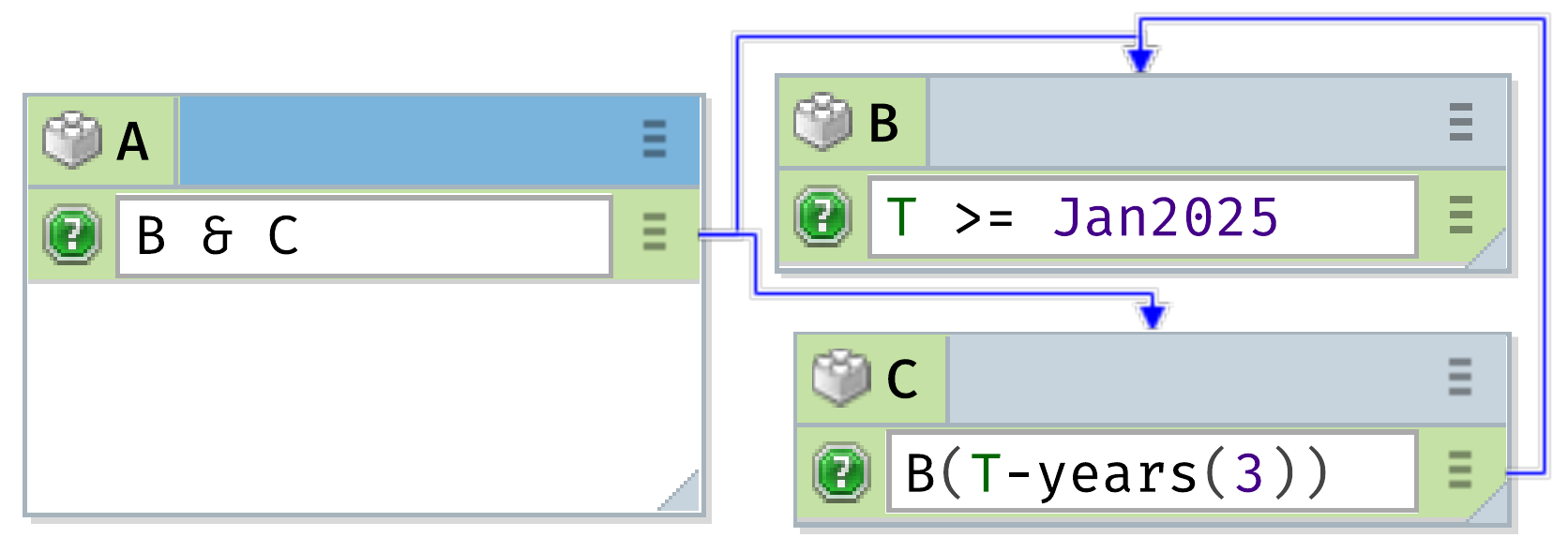}
        \caption{A trivial example of modeling time-dependency}
        \label{fig:trivialExample}
\end{figure}

Prospective roadmapping is an activity that looks into the future. Consequently, the model at a certain point in time, not at the present point, is of interest. A trivial approach that could be used with existing modeling tools is shown in Figure~\ref{fig:trivialExample}: the availability of block \irisinline{A} depends on the availability of \irisinline{B} and \irisinline{C}. \irisinline{B} is available if the time $T$ is larger than or equal to January 2025. \texttt{C} is available 3 years after the availability of \texttt{B}. Using this annotation, a user would have to manually evaluate the emerging equation $(T \geq Jan2025) \land (T - years(3) \geq Jan2025) \implies avail(A)$ for every point of interest. Although this is easy to solve for this small example, it becomes rather difficult with increasing complexity of a model such as our running example \textit{Fuse}. For instance, the fuse's availability depends on the vehicle's \irisinline{TotalCurrent} which in turn is the sum of all currents of all consumers inside the vehicle (cf. Figure~\ref{fig:runingExample}). But the current of the central processing units changes over time due to the linear growth of \texttt{TFLOPS} in the block \texttt{DetectionSoftware}..

\begin{figure}[htb]
        \centering
        \includegraphics[width=\linewidth]{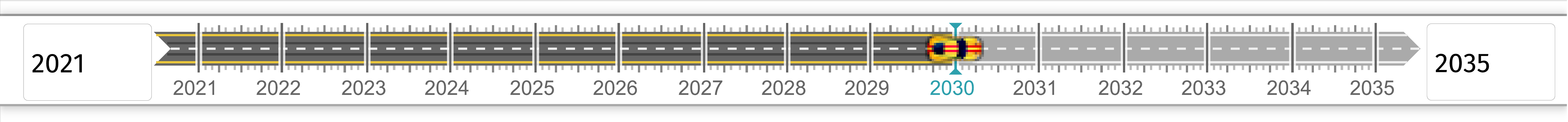}
        \caption{Time slider to view model at certain point of time in the future}
        \label{fig:timeslider}
\end{figure}

To address this complexity challenge, we introduce a time slider as shown in Figure \ref{fig:timeslider}. The time slider enables users to change $T$ by dragging a handle (which looks like a car in our prototype) to a certain point of time. After the user has set $T$, \TextIris\ evaluates the equation system and displays the result.

\subsection{Availability Highlighting}

Simply hiding or showing a block depending on its availability does enable roadmap engineers to analyze time-dependency, availability, and changes over time. Therefore, to indicate time-dependency, we change the style of an element accordingly. We identified six different cases of availability that are relevant to roadmap engineers:

\subsubsection{Always-available}
If a block has no explicit requirements, we assume that it is implicitly always available. This can be regarded as a kind of baseline. Consequently, there is no need to highlight  always available blocks in a special manner. As shown in Figure~\ref{fig:runingExample}, the block \irisinline{Headlights} is always available.

\subsubsection{Currently-available}
\begin{figure}[htb]
    \centering
    \includegraphics[width=0.8\linewidth]{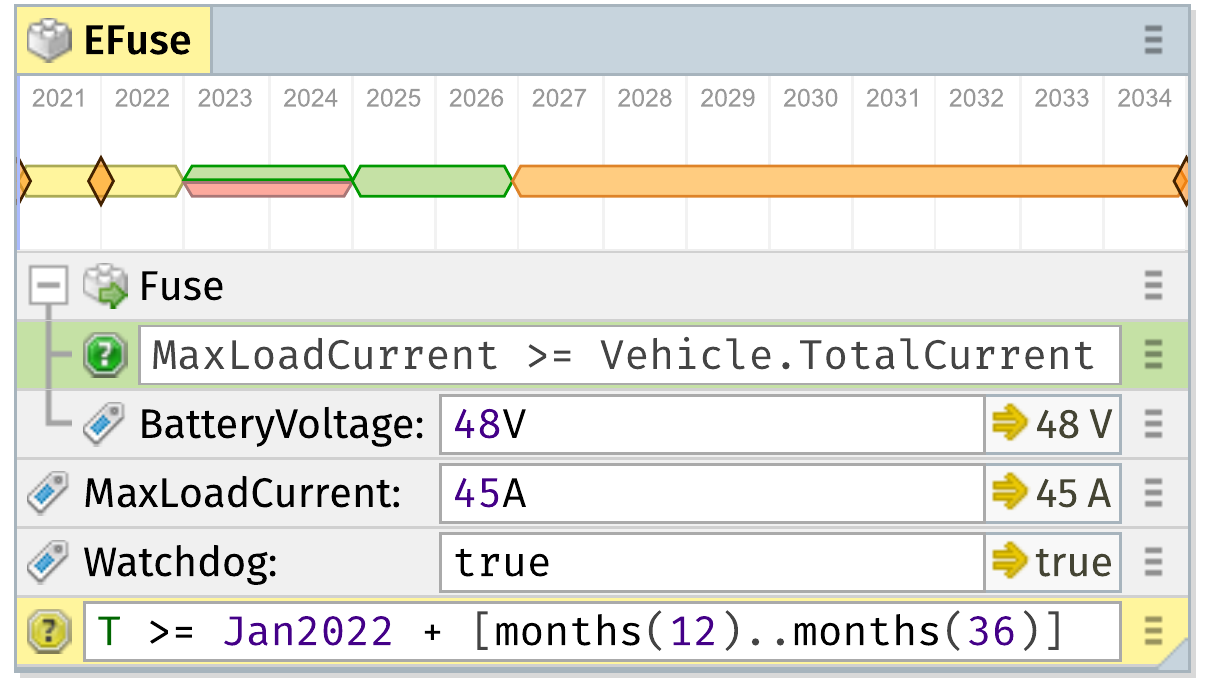}
    \caption{Block \texttt{EFuse} is not yet available ($T=Jan2021$) but will be available in the future.}
    \label{fig:Efuseimpl}
\end{figure}

In contrast to \textit{always-available}, if a roadmap engineer has explicitly modeled at least one requirement, a block becomes available, if all its requirements are evaluated to \textit{true} at the current point in time. In Figure~\ref{fig:Efuseimpl}, block \irisinline{EFuse} contains two requirements. The first one, \irisinline{MaxLoadCurrent >= Vehicle.TotalCurrent}, is \textit{true} or satisfied as indicated by a green background color. However, \irisinline{EFuse} is not available because of the second requirement, as we will discuss in the next case.

\subsubsection{Not-yet-available} A requirement is currently not satisfied but will become satisfied in the future. We highlight this kind of requirement in yellow such that a roadmap engineer can identify the reason why a requirement is still not available, when it will become available, and why it is not yet available. This knowledge can be used to focus on specific research and development to speed up a new innovation.

Back to Figure~\ref{fig:Efuseimpl}, for the currently selected $T=Jan2021$ the second requirement \irisinline{T >= Jan2022 + [months(12)..months(36)]} is evaluated to \textit{false} which in turn leads to non-availability of \irisinline{EFuse}. Changing $T$ to be greater or equal than January 2023 (being the date 12 months after January 2022) would \textit{maybe} satisfy the requirement and thus the availability of \irisinline{EFuse}. Changing $T$ to a value greater or equal than January 2025 satisfies the requirement and thus the availability of \irisinline{EFuse}.
This is because \irisinline{T >= Jan2022 + [months(12)..months(36)]} is \emph{true} for any $T \geq Jan2025$.

\subsubsection{No-longer-available}
A requirement that has been satisfied before is no longer satisfied at the currently select point of time. This can be seen in Figure~\ref{fig:runingExample}: \irisinline{EFuse}'s \irisinline{MaxLoadCurrent} is no longer greater or equal than the vehicle's total current and is highlighted in orange.

\subsubsection{Maybe-available}

A requirement is maybe satisfied if the value of its expression evaluates to \textit{maybe}. This is the case for the second requirement presented in Figure~\ref{fig:Efuseimpl} and any $T$ in between January 2023 (\irisinline{Jan2023}, inclusive) and January 2025 (\irisinline{Jan2025}, exclusive),
because the right hand side of the relation \irisinline{T >= Jan2022 + [months(12)..months(36)]} evaluates to \irisinline{[Jan2023..Jan2025]} (see Subsection~\ref{subsec:interval-arithmetic}), yielding the relation \irisinline{T >= [Jan2023..Jan2025]}.

\subsubsection{Never-available}
A block also can be never available. We use red color to highlight not satisfiable requirements to help roadmap engineers to identify required innovation.

\begin{figure}[htb]
    \centering
    \includegraphics[width=\linewidth]{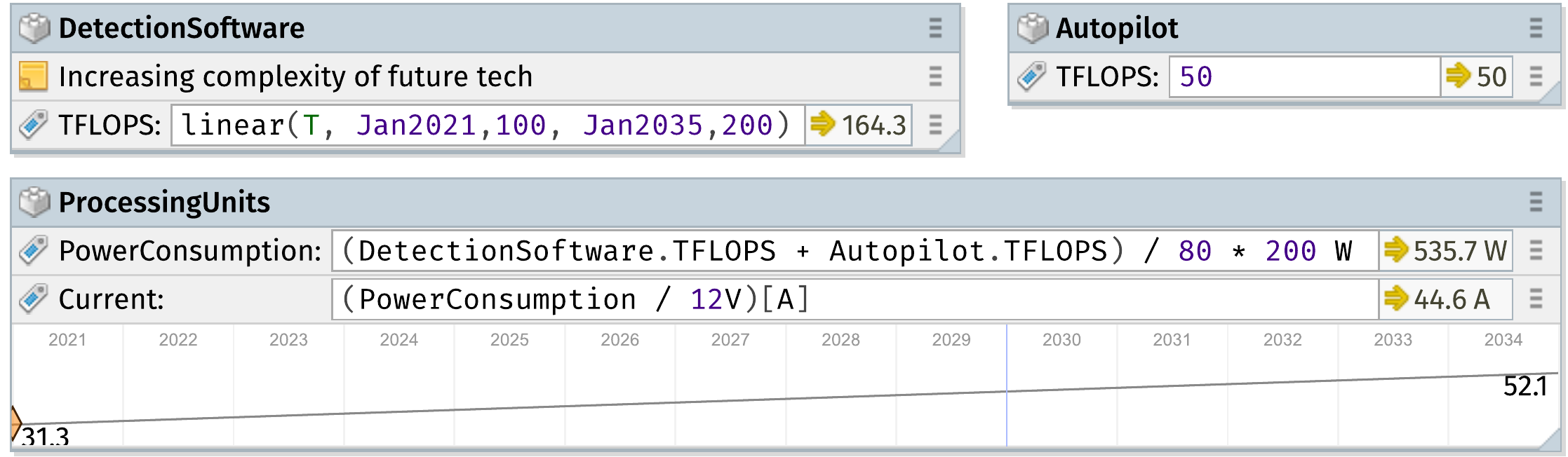}
    \caption{\texttt{Processing Units}' current increases over time ($T=Jan2030$) because of linearly growing \texttt{TFLOPS} of \texttt{DetectionSoftware}.}
    \label{fig:propertyOverTime}
\end{figure}
\subsection{Time-dependency Highlighting}

There are more aspects of time-dependency we did not cover so far. As already introduced, the value of a property can be time-dependent in the sense of changing over time. For example, as shown in Figure~\ref{fig:propertyOverTime}, the \irisinline{Current} of the processing units increases linearly from \SI{31.3}{\A} in 2021 to \SI{52.1}{\A} in 2035 because it depends on the growing processing power in \texttt{TFLOPS} of \texttt{DetectionSoftware} modeled via formula \irisinline{linear(T, Jan2021,100, Jan2035,200)}. The result of a formula for current $T$ is always displayed next to a formula. Apart from the current value the change of a property's value over time is also important for roadmap engineers. Therefore, we plot a property's value curve. Figure~\ref{fig:propertyOverTime} shows the curve of the increasing \irisinline{ProcessingUnits}' current.

Furthermore, we also plot satisfiability of blocks and requirements over time. Figure~\ref{fig:Efuseimpl} shows the availability of \irisinline{EFuse}. The plot shows that \irisinline{EFuse} is not-yet-available for $T < Jan2023$, maybe available for a period of two years, and  becomes available in 2025 but will be come unavailable again from 2027. This is the moment in time where the vehicle's \texttt{TotalCurrent} becomes greater than \irisinline{EFuse}'s \irisinline{MaxLoadCurrent}, because of the increasing \irisinline{Current} of the processing units.

\begin{figure}[htb]
    \centering
    \includegraphics[width=\linewidth]{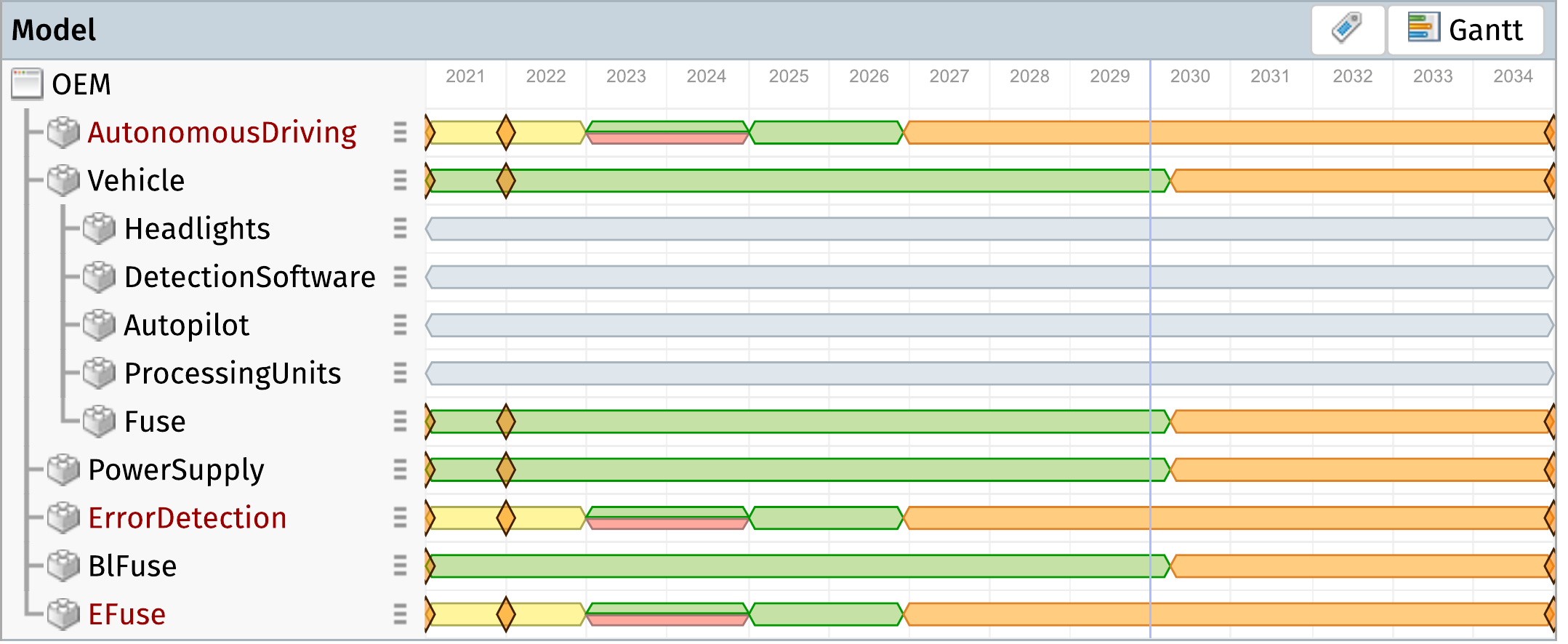}
    \caption{Overview of the availability of all blocks over time.}
    \label{fig:gantt}
\end{figure}

In order to further inspect the availability of blocks, we also provide a summarizing overview, as shown in Figure~\ref{fig:gantt}. Here, the availability of each block is displayed as a sequence of colored bars according to its availability. A green bar indicates the case \emph{currently-available}, with a light blue indicating the sub-case \emph{always-available}, whereas a yellow bar indicates the case \emph{not-yet-available} and an orange bar the case \emph{no-longer-available}. The half-green and half-red bar indicates the case \emph{maybe-available}, the case \emph{never-available} is not shown in this figure but would be visualized by a red bar.

\subsection{Chosen technology framework}\label{sec-visualization-technology}

In the following, we sketch how we technically realized the presented language, the solving of the constraint system as well as the visualizations of the graphical models as well as the analysis results. 

Our work was conducted in a collaborative research project with industrial partners from the embedded systems domain. They stated two key requirements in order to ensure applicability in industrial practice: (1) The system had to be web-based in order to avoid any local installation and ensure an easy central upgrade process, and offer a (2) high usability for industrial domain experts. Both requirements are in our opinion not met by today's frameworks and technologies, e.g., the EMF-ecosystem, for the development of industrial-grade applications.

With respect to web-based systems, technologies like Sirius Web \footnote{\url{https://www.eclipse.org/sirius/sirius-web.html}} were not yet available when we started our research in 2018. With respect to usability, our personal experience stemming from nearly two decades of developing domain-specific modeling languages and tools (e.g.~\cite{DBLP:journals/sttt/BurmesterGNTWWWZ04,DBLP:conf/dagstuhl/PriesterjahnTHHS07,DBLP:conf/sle/MaroSATG15,DBLP:conf/gg/0001BGGKOT17,DBLP:conf/staf/TichyMJH20}) is that there is an inherent trade-off between productivity gained by using off-the-shelf modeling technology frameworks and flexibility gained by developing a modeling environment for a specific domain-specific language from scratch using the much broader programming language ecosystem. When starting the development, we chose the JavaScript/\allowbreak TypeScript-ecosystem\footnote{\url{https://www.typescriptlang.org/}} with the technologies Node.js\footnote{\url{https://nodejs.org/}} and React\footnote{\url{https://reactjs.org/}} to develop the tool as we valued the flexibility for our concrete case higher than the productivity gain of modeling technology frameworks. For example, this flexibility allowed us to easily visualize results from the solving inside the concrete graphical syntax as seen in Figures~\ref{fig:Efuseimpl} and~\ref{fig:propertyOverTime}, support collaboration features in \TextIris\ (naturally embedded into the architecture of Flux ~\cite{Pietron+2021}), as well as easily integrate external systems from our industrial partners via REST-APIs.

Nevertheless, we use the standard ingredients of modeling language engineering: meta-models, models, model transformations, and a concrete graphical syntax. Our meta-model from Section~\ref{ssc:modeling-language} is defined using TypeScript interfaces. An important aspect here is that all attributes are read-only as our models (as JavaScript objects conforming to the interfaces) are immutable and form persistent data-structures. Hence, all user changes result in \enquote{new} immutable models while applying substructure sharing to minimize the memory impact. The immutability ensures that all changes are done centrally in change actions and not in arbitrary places throughout the code.

Our symbolic transformations are similar to in-place transformations, endogeneous model transformations that check a pre-condition on a specific part of the model and return an updated model part. Comparing these transformations to our own graph transformation framework Henshin~\cite{DBLP:conf/gg/0001BGGKOT17}, it is more difficult to match complex graphs in the pre-condition, however, mathematical calculations (which many of our symbolic transformations deal with) are much easier to express.

Finally, our React-based frontend follows the projectional-editing approach based on the Flux pattern~\cite{Gackenheimer2015} which blends well with the aforementioned immutable models. The parsing of mathematical expressions is realized by a recursive descent parser with subsequent stages for name resolving and type inference. The resulting syntax trees are represented with immutable algebraic data types, which simplifies the implementation of symbolic transformations and improves testability.

In summary, our experiences with the chosen technologies for our specific context have been generally good and we enjoy the gained flexibility and the power of the ecosystem, both in terms of development infrastructure as well as libraries.

\section{Evaluation}\label{sec:evaluation}

The purpose of our conducted evaluation is to
examine the applicability of \TextIris\ to a real-world innovation case by involving domain experts. This involves assessing the suitability of \TextIris\ for the realization of tasks regarding the modeling and analysis of innovations as well as identifying general potential for improvement. Thus, we ultimately examine \TextIris\ with respect to an inherent characteristic of second generation technology roadmaps, namely that, according to Letaba et al.~\cite{Letaba.2015}, they enable the comparison between current technologies and potential innovations. Therefore, first we investigate whether the concepts \TextIris\ are fundamentally suitable to model aspects of the development of an innovation. Second, we investigate how domain experts perceive the usefulness of our modeling approach.
To achieve this goals, this section first describes our research questions and case study design, the materials used, the execution, threats to validity, and the results of the evaluation.

\subsection[Research Questions and Study Design]{Research Questions and Study Design}
\label{sec:study_design}

In order to clarify the objectives of our evaluation, we pose the following \textbf{research questions}:

\begin{enumerate}
    \item[RQ1] Does the Domain-Specific Language (DSL) of \TextIris\ support modeling and analyzing relevant aspects in the development of an innovation?
    \item[RQ2] How do domain experts perceive the support of \TextIris\ in modeling and assessing potential innovations?
\end{enumerate}

With research question RQ1, we investigate to what extent \TextIris\ offers the possibility to capture and analyze the information necessary for the development and evaluation of an innovation. This includes, for example, the structural design of involved components, their dependencies, requirements, and properties.
The aim is to identify weaknesses with regard to the maturity of the meta-model in \TextIris\ as well as its analysis capabilities.
Research question RQ2 focuses on exploring the opinion of domain experts regarding the suitability of \TextIris\ for developing innovations.
This includes an adequate and comprehensible representation of relevant information for domain experts. For this, qualitative feedback from domain experts plays a crucial role, as they are best able to evaluate how \TextIris\ compares to previously used roadmapping tools. From this, it can be deduced whether they consider our approach for technology roadmaps to be useful, which would improve its chances for adoption in industrial practice.

Our \textbf{methodological approach} to answering these research questions is reflected in our design and execution of the evaluation. We decided to implement the evaluation as a \textit{case study}, as this is very well suited for the industrial evaluation of software engineering methods and tools~\cite{Runeson2012}.
It enables an interactive, iterative and flexible approach, which we implemented in the form of a guided expert evaluation.
For this reason, we formed \textit{focus groups}~\cite{Singer2008} from different levels along the automotive value chain, i.e., OEM, Tier~1, and semiconductor manufacturers~(SCMs), for the purpose of data collection. Employees from three different companies participated in these focus groups. The task suitability of the \TextIris\ tool and missing concepts were discussed during these focus groups with the participating domain experts using a real-world example from industry. For this purpose, we used historical data from the prototypical development of a smart sensing fuse (cf. running example described in Section~\ref{sec-example}), from which we extracted the information about the innovation and the different considered alternative solutions as well as the used innovation process. These two aspects are described in more detail in the following subsection.

In our conducted guided expert evaluation, the following four roles were involved:
The operation is taken over by a (1)~\textit{tool operator} who was involved in the development of the tool. We consider this to be appropriate because, on the one hand, the evaluation of the usability of the \TextIris\ tool was not the main focus of the study and, on the other hand, tool training would have meant additional effort for the domain experts from industry, for which the various participants could not provide resources to the same extent. In addition, due to the fact that the tool is operated directly by the tool developer, any usability difficulties that may arise in the operation of the tools do not cause irritation among the domain experts or distort the results of the actual measurement objectives~-- the task suitability of the tool, the assessment of the implemented concepts as well as the identification of missing concepts.

In addition to the role of the tool operator, our study design also includes the role of (2)~\textit{moderator} and (3)~\textit{minute taker}. Hence, the task of the moderator is to guide the evaluation and ask purposeful questions that are conducive to achieving the evaluation goal. For this purpose, the moderator has intensively studied the demonstrator beforehand in order to stimulate the thought process about the demonstrator among the domain experts during the workshops. While conducting the study, participants are asked to express their thoughts and comments regarding positively or negatively perceived aspects regarding the \TextIris\ tool. The minute taker has the task of recording the insights gained during the evaluation and the statements made by the domain experts. Furthermore, notes on technical aspects are taken, e.g., how well a specific aspect could be handled by \TextIris.

And finally, the participants from the automotive industry attending the focus group meetings took on the role of (4)~\textit{domain experts}.

\subsection{Study Material}
\label{sec:study_material}

\begin{figure}[!h]
    \centering
    \includegraphics[width=\textwidth]{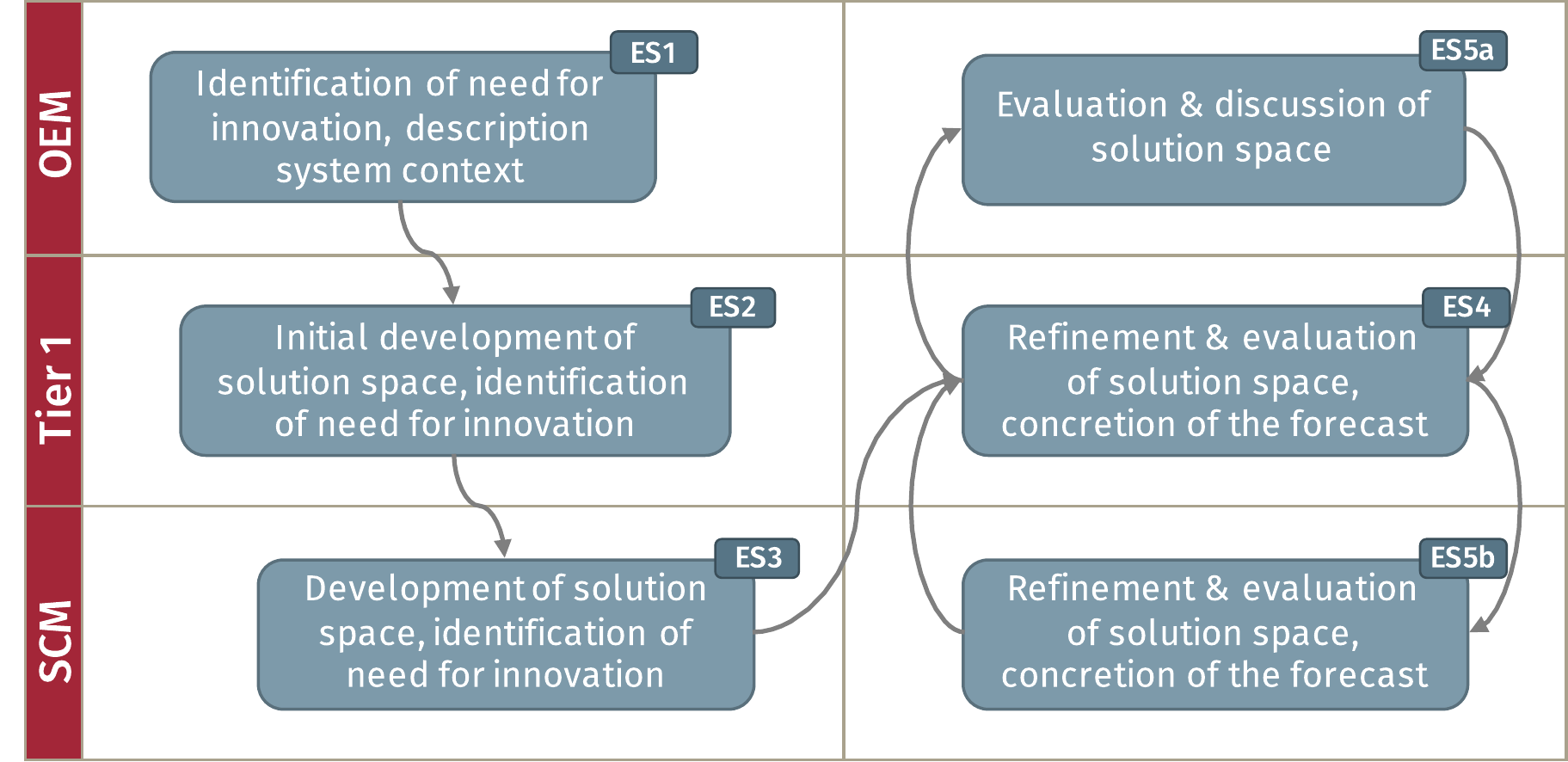}
    \caption{Derived process of the past \textit{smart sensing fuse} development}
    \label{fig:process}
\end{figure}

Since it is our goal to evaluate the suitability of our tool and the realized concepts in terms of supporting the identification of the need for innovation or still not yet existing technical solutions, we choose the smart sensing fuse as our object of study (cf. Section~\ref{sec-example}). To set up a realistic evaluation setting and utilize a suitable modeling artifact, we collected information about the past technical innovation process of the smart sensing fuse with a focus on required innovations, involved stakeholders along the automotive value chain, and challenges that arose during the development process. Therefore, we conducted multiple workshops with our partners from the automotive industry before the actual evaluation. As a result, we are able to identify involved value chain stakeholders as well as a set of evaluation scenarios (cf. Figure~\ref{fig:process}) that reflect the historical innovation development process and thus form the basis of our case study. Furthermore, we composed a collection of materials containing context and technical details that each study participant received before the conduction of the study.

\textbf{Involved Value Chain Stakeholders.}
The idea for the smart sensing fuse was developed by a joint work of an OEM and a Tier~1 and mainly driven by the latter one. Since a smart sensing fuse relies on a semiconductor, during the technical innovation process also SCMs  contributed. In consequence, we require representatives from each type of manufacturer.

\textbf{Evaluation Scenarios.}
Based on the collected information about the technical innovation process, we derived six evaluation scenarios (ESs) shown in Figure~\ref{fig:process}. An ES should simulate an excerpt of the whole technical innovation process addressing specific goals. For each ES we identified and listed the value chain partners involved, technical as well as content objectives, concrete tasks that should be completed within the ES, required artifacts (e.g., models created in a previous ES) which should serve as input, and the expected output of an ES, see Table~\ref{tab:es}. While content objectives describe which part of the technical innovation process should be done, technical objectives focus on features of \TextIris\ that should be used especially.

\begin{table*}[]
{\footnotesize
\begin{tabularx}{\linewidth}{@{}p{0.26cm}*3{>{\centering\arraybackslash}p{0.13cm}}XX@{}}
\toprule
                            & \rotatebox{90}{\textbf{OEM}} & \rotatebox{90}{\textbf{Tier 1}} & \rotatebox{90}{\textbf{SCM}} & \textbf{Content Objectives}                                                                                                                                                                                                                          & \textbf{Technical Objectives}                                                                                                                                                                                                                                          \\ \midrule
\textbf{ES1}    &   \checkmark  &        &                                                                                  & Based on future trends such as reducing a vehicle's CO$_2$ emission or save installation space, the OEM should derive the need for the innovation \enquote{smart sensing fuse}.           & The OEM should systematically and in a structured way capture the need for an innovation, describe the system context, and derive requirements to the Tier 1.                                                                                  \\
\textbf{ES2}    &     &    \checkmark    &                                              & The Tier 1 develops different solution concepts for a smart sensing fuse. At that point of time, available semiconductors are not able to handle currents which are required by that specific car application.         & The Tier 1 should import the initial requirements by the OEM to \TextIris, define an initial solution space, and identify the need for an innovation, and derive requirements to the SCM.                                                          \\
\textbf{ES3}   &     &        &  \checkmark   & The SCM constructs different approaches that could satisfy the requirements of the Tier 1.      & The SCM should import the initial requirements by the Tier 1 to \TextIris, define an initial solution space, and develop a first idea how to address the need for innovation. The SCM's broad idea should be exported and sent back to the Tier 1.                                                                                                                                                             \\
\textbf{ES4}    &     &    \checkmark    &            & Now, the Tier 1 is able to develop three concrete solutions (a discrete, a partially integrated and a fully integrated circuit). Furthermore, possibly added values by the smart sensing fuse compared to a blade fuse should be part of the KPI benchmark.  & Based on the idea for an innovation by the SCM, the Tier 1 should evolve its solution space. Solution options should be refined and benchmarked on the basis of internal KPIs.                                                         \\
\textbf{ES5 a/b} &  \checkmark   &   \checkmark     &  \checkmark      & By consulting with the OEM and the SCM the Tier 1 is able to develop application-specific solutions. Possible problems and challenges should be identified as early as possible.              & Tier 1 discusses, refines, and evolves together with the OEM (5a) and the SCM (5b) the developed innovation. The functionality of \TextIris\ to analyze a model should support \enquote{what-if} discussions or discussions of alternatives in general.                                                                    \\ \bottomrule
\end{tabularx}}
\caption{Summary of involved stakeholders, content, as well as technical objectives for each evaluation scenario}
\label{tab:es}
\end{table*}

We intend that from ES1 to ES5 the model of the smart sensing fuse should be iteratively evolved. For this reason, at the end of each ES, the resulting model should be exported and should serve as an input for the subsequent ES. Figure~\ref{fig:process} illustrates the derived ES and their order of execution. Each blue box represents and briefly summarizes the activity of an ES. An arrow between two ES represents a communication and information flow. ES1 to ES3 mainly focus the initial modeling of a solution space and requirements per manufacturer. The information flow can be described as top-down. In each ES mentioned before, only a single manufacturer is involved. In contrast, in ES4 and \mbox{ES5a/b} the communication flow is cyclic. Now, the OEM and the Tier~1 (ES5a) or the Tier~1 and SCM (ES5b) discuss and evolve the model together within the same session. Especially in \mbox{ES5a/b}, the features of \TextIris\ to analyze a model and support \enquote{what-if-analyses} should be used.

\subsection{Execution}

The execution of our study was based on the identified ES for the smart sensing fuse (cf. Figure~\ref{fig:process}). We were able to recruit five domain experts from automotive industry: 1)~From the OEM three domain experts that have relevant knowledge but were not involved in the past technical innovation process participated. 2)~From the Tier~1, the main engineer of the smart sensing fuse participated. 3)~From the SCM, an engineer participated who is familiar with the domain, as well as cross-sectional as a development process methodologist.

Each ES was addressed in a separate session, which was held virtually via a video conference system. Every session was attended by the moderator, the minute taker, the tool operator for \TextIris, and the domain experts from one or two companies regarding Table~\ref{tab:es}.

The study was conducted over a period of two weeks in ten sessions in August~2020. Each session lasted between 1 and 2.5 hours. Before every session, we handed out the collection of material describing the smart sensing fuse, a short video describing the features of \TextIris\ from a user's perspective, and the textual description of the evaluation scenario (cf. Table~\ref{tab:es}) to the participants.

\subsection{Threats to Validity}

Our study reveals some threats to validity, which we explain below.
A threat regarding \textit{construct validity} is a different level of knowledge of the respective participants (i.e., the domain experts) concerning the considered smart sensing fuse. We tried to reduce this risk by providing all participants with information about the smart sensing fuse and its historical innovation process before the respective sessions. Nevertheless, insufficient preparation of the participants, which takes place individually before the respective sessions, cannot be completely excluded. Moreover, there is a threat that the case under consideration does not correspond to real practice. We have countered this by selecting the smart sensing fuse as a representative and real example from the automotive industry and reconstructing its innovation process together with experts in the field to conduct our study. Furthermore, we avoided a biased view or evaluation of our tool by recruiting different domain experts along the value chain for our study. The lack of prior knowledge with the tool by the domain experts may also have a negative impact the study. In order to counter this threat, we opted to utilize expert tool users, such that the domain experts can focus on the modeling itself.

The \textit{internal validity} is limited by the predefined phases of the innovation process, which was based on historical data. In this respect, there is an influence of the participants by the provided material but also by the guided focus groups, since it is conceivable that participants relied too much on these guidelines and thus possibly inhibited the creation of new situations. In contrast, the preparation of the historical innovation process offered the opportunity to discuss and reflect on real problems that had arisen. In addition, we cannot exclude the possibility that the focus group meetings in the form of video conferences may pose a threat. While in a face-to-face meeting it might be more obvious for the moderator whether a participant wants to comment on a certain aspect, a non-verbal communication is sometimes difficult to recognize via video conference. Moreover, influences on the results during the analysis were reduced by having two independent researchers evaluate the collected data and combine them into problem classes, which were then discussed.

Case studies generally have a low \textit{external validity} due to their focus on individual cases. Hence, there exist some threats in this regard. With five domain experts, we have only a small number of participants, which is not necessarily representative. Nonetheless, despite the difficult availability of domain experts from industry, we still managed to get domain experts from three different companies representing all roles along the value chain to participate in our study. In this respect, it should be noted that the majority of the participants did not develop the smart sensing fuse themselves and were therefore required to make ad-hoc considerations about its requirements during the sessions. The advantage of these participants is that they are not influenced by the historically past events and are therefore not biased in their thinking.

A threat to the \textit{reliability} of our study is posed by the questions asked by the moderator during the sessions. Since these questions were not prepared, but arose from the respective situation and the sessions themselves could not be recorded due to confidentiality restrictions on the part of the industry participants, it would not be possible to reproduce the sessions identically. The minutes only contain the identified comments of the experts, without any judgement of the minutes-taker. However, the reproducibility of the extracted results is guaranteed, since the minutes were used as a basis for this.

\subsection{Results}
\label{sec:evaluation_results}

Based on the minutes recorded during the sessions by the minute taker, two independent researchers analyzed the data and discussed their results. Thereafter, the researchers structured the identified problems and potential improvements mentioned by the domain experts. In a joint meeting, the collected results were consolidated.
In this section, we present the results of the evaluation as follows: 1)~We describe the result artifacts, i.e., the developed models (cf. Section~\ref{sec:models}). 2)~We present identified technical findings that form the basis for answering research question RQ1 (cf. Section~\ref{sec:technical_findings}). 3)~Furthermore, we present the qualitative feedback from the domain experts about the usefulness of our approach w.r.t.~research question RQ2 (cf. Section~\ref{sec:qualitative_feedback}).

\subsubsection{\TextIris\ Models}
\label{sec:models}

Within the evaluation scenarios~1 to~5 (see Section~\ref{sec:study_material}, Figure~\ref{fig:process}), different models were created specifying relevant information and representing the solution space exploration at the OEM, Tier~1 and SCM levels. In total, the models consist of about 349 elements, where 43 are components, 211 are properties, 52 are requirements, 31 are notes, and the remaining elements are distributed among the other element types.

On the one hand, the models created during the evaluation represent the OEM's requirements for a new type of smart sensing fuse for various electrical consumers, such as starter, rear headlamp, gear oil pump, sound system, front loader, etc..
On the other hand, these requirements for the EFuse are forwarded to the Tier~1, so that the latter can build up two solution spaces~-- implementation by means of an integrated highside switch and a MOSFET plus ASIC. For the implementation of a switching element with protection function and reverse protection, Tier~1 set up a benchmarking with regard to three concrete solutions available on the semiconductor market. This benchmarking from Tier~1 was based on solution spaces that the SCM communicated to the Tier~1 including two smart power switches and one integrated gate driver.
These alternative solutions were compared based on a total of 19 properties (e.g., over temperature protection, surface, rated current, switching time, time availability) and four basic requirements (reverse protection, cold crank pulse duration, big capacity load, intrinsic protection), of which only one solution adequately met all needs.

\subsubsection[Technical Findings regarding the implemented metamodel (RQ1)]{Technical and Conceptual Findings Regarding the Implemented Metamodel (RQ1)}
\label{sec:technical_findings}

Overall, the \TextIris\ tool proved to be robust during the entire evaluation, so that no system crashes occurred. The created models could be saved and reloaded without loss of information. Nevertheless, during the evaluation as well as during a preceding intensive test phase, a total of \textbf{82 technical findings} were uncovered and most of them were resolved before the evaluation began. These included, for example, deficiencies with regard to interaction with the user interface, such as when moving blocks, display errors, problems with key combinations, when scrolling, when entering spaces or when zooming.

Basically, during the evaluation sessions, the various information of the EFuse historical demonstrator could be captured and presented at different levels of the value chain using \TextIris. This mainly includes the definition of components, their interrelationships as well as the abstract textual specification of requirements from the OEM side, such as \textit{\enquote{can reconnect loads}} or \textit{\enquote{can measure current flow}}. We would like to emphasize at this point that \TextIris\ does not claim to provide the full functionality of requirement management tools, as this is not decisive with regard to feasibility analyses in the early phases of an innovation's development.
Furthermore, especially on the side of Tier 1, properties could be defined that further specify the requirements, such as \textit{\enquote{big capacity load >= \SI{10}{\milli\farad}}} or \textit{\enquote{cold crank pulse duration <= \SI{19}{\milli\second}}}. \TextIris\ also enabled calculations to be performed, such as the surface area and
the required cable cross sections.
Hence, requirements and properties could be defined and their fulfillment determined using the \TextIris\ internal solver. In addition, \TextIris\ made it possible to standardize redundant information through the use of inheritance relationships and thus reduce the scope of the model.

In addition, the participation of industry partners during the evaluation also resulted in the identification of some complementary \textbf{conceptual findings}.
This includes, for example, a previously missing \textbf{possibility to specify conditional existences of blocks or implications}.
For example, a discrete MOSFET requires temperature monitoring as opposed to an integrated solution. This implies a backward communication of requirements from the SCM to Tier~1 regarding the need to ensure temperature monitoring. This means that the existence of certain components implies the existence of other components, as in this example of a temperature sensor.
Another example in the context of the EFuse is the material of the EFuse's housing. For better heat dissipation, the Tier~1 may install a housing made of aluminum.
However, this implies that the OEM must also ensure that elements that realize heat exchange are included in the design of the system. This requires the possibility of bidirectional communication of requirements and conditions of existence between Tier~1 and OEM.

Moreover, \textbf{change management mechanisms} are currently missing in \TextIris. This means that exporting and updating along the value chain is currently technically possible, but there are no mechanisms to make changes traceable for the various partners involved and to inform them of a change that has been made. We observed this during the evaluation because it was difficult for different partners along the value chain to figure out which elements were changed, added, or removed after re-importing a model.
This reveals that change detection and difference visualization mechanisms are needed and also desired by domain experts.

Furthermore, in the context of solution space evaluation, it was considered interesting by the participants in the evaluation to be able to \textbf{assign a confidence level to the values and value ranges} communicated by a partner. It should be possible to actively communicate how sure the respective modeler or domain expert is about the specified value. Is it a hard fact or a rough estimate.
Moreover, the domain experts considered a \textbf{priority level for key performance indicators (KPIs)} to be useful. Currently, \TextIris\ lacks a possibility to weight internal KPIs so that a solver can automatically select the best solution evaluated according to individual KPIs. Regarding this, the importance of KPIs must also be taken into account, because not all expected properties are equally important. Therefore, a possibility to classify KPIs, such as \enquote{Must be fulfilled}, \enquote{Possibly fulfilled}, \enquote{Nice to have}, is missing in the context of the solution space evaluation.

Based on the above findings presented, we can derive an answer to research question RQ1 in the following. In the evaluation, we were able to map relevant aspects of the development of an innovation and therefore we believe that \TextIris\ fundamentally supports the modeling of innovations. This means that the concepts provided in \TextIris, such as components, their interrelationships, inheritance hierarchies, properties and requirements, provide a good basis for modeling innovations. Therefore, the \TextIris\ metamodel was sufficient for modeling these aspects, as it was possible to model the thoughts of the domain experts together with the tool operator using the DSL of \TextIris\ during the evaluation sessions. Nevertheless, there is potential for extending the concepts already implemented (e.g., by adding change management mechanisms or confidence levels).
It should also be noted, however, that it would be quite necessary for domain experts to undergo training in the usage of \TextIris\ beforehand, so that they would be able to create such models on their own.

\subsubsection[Qualitative Feedback from the domain experts (RQ2)]{Qualitative Feedback From the Domain Experts (RQ2)}
\label{sec:qualitative_feedback}

The qualitative feedback on \TextIris\ from the domain experts is generally positive. Compared to MS Excel or MS PowerPoint, which are usually used for representing roadmaps in an industrial context, as reported by the participating domain experts, \textit{\enquote{\TextIris\ offers [the domain experts] the advantage of being able to represent relationships more quickly, since there are predefined elements such as ready to use blocks with a well-defined semantic}}. \TextIris\ also \textit{\enquote{enables the presentation of complex facts and relationships, which provide information that can also be subjected to respective benchmarking}}. Furthermore, the domain experts appreciate that the application of the \TextIris\ tool opens up the possibility of considering several solution spaces in parallel which is not the case in current practice.
As domain experts tell us, the solution spaces are too large or there are too many solutions to consider in detail, considering the techniques available today.
In contrast, the participating domain experts assume that solution spaces can be constructed \textit{\enquote{more easily and quickly with \TextIris}}. Thus, \TextIris\ offers the possibility to evaluate even complex systems and to consider edge solutions, which in turn can lead to insights that would not have been considered in today's context. For the domain experts, it was possible to perform analyses based on the data received from other partners and thus to benchmark alternative solutions as well (in particular benchmarking on the part of Tier~1). This made it possible to compare them clearly.

Moreover, collaboration between the industry partners and the exchange of information between these partners along the value chain was possible, at least to a limited extent. However, the participants noted that in a real-world scenario exporting a whole model is not compliant and might reveal intellectual property. Thus, a way to export only certain aspects of a model or export some kind of a \textit{black box} needs to be developed. This feedback from domain experts is related to the addressed need for implementing change management mechanisms, as already described in the previous section (see Section~\ref{sec:technical_findings}).

We think that these statements of the domain experts represent an answer to research question RQ2. The domain experts give \TextIris\ positive feedback compared to the tools used so far, especially regarding the analysis of solution spaces. The basis for this is also the good comprehensibility of the models and the elements they contain. The domain experts also positively emphasized the good comprehensibility of the composition of calculations, whose highlighting (cf. Figure~\ref{fig:referenceHighlighting}) is based on the familiar behavior of Excel. This enabled the domain experts to quickly find their way around.

In summary, the evaluation showed that \TextIris\ enabled the various contributions of the partners along the value chain to be captured separately and in a structured manner. For this purpose, the modeling language implemented in \TextIris\ and derived from the meta-model turned out to be suitable for capturing the thought processes of the domain experts in a structured manner. The system structure, which is relevant for the development of an smart sensing fuse, could be modeled with all innovation-relevant components, requirements, and properties according to the underlying historical demonstrator using \TextIris. It was possible to use measurement units and formulas and to define solution spaces at the various levels along the value chain, which is also reflected in the positive feedback from the domain experts.

\section{Conclusion and Future Work}\label{sec-conclusions}

In this paper, we presented a model-based approach based on a domain-specific modeling language for the creation of technology roadmaps as well as a corresponding interactive user interface. 
The modeling language supports time-dependent properties and the time-dependent availability of structural elements and thus allows the description of a range of various valid models over time. In addition, the language supports interval arithmetic and ternary logic to model uncertainty, such as uncertainties of the domain experts which are expressed in inaccurate values, as they usually occur in the course of an innovation process. The solver integrated into \TextIris\ solves the global system of equations spanned by properties, requirements, and KPIs using repeated symbolic transformations enabling the use of traceability information in the visualization.
In addition, the realized visualization and interaction concepts in \TextIris\ support the evaluation of time dependencies within properties and structural components of a technology roadmap. This allows roadmap engineers to easily substantiate technology decisions and apply updates to the technology roadmap.

Furthermore, we have shown in the evaluation that the modeling language and the visualizations in the \TextIris\ tool are suitable to capturing the thought processes of industrial domain experts during an innovation process in a structured and understandable manner. It turned out that the visualizations and interaction concepts implemented in \TextIris\ are intuitively understandable and applicable for roadmap engineers, who are domain experts but not modeling experts.

However, the modeling itself has not been performed by domain experts themselves, which necessitates additional empirical work. 
In the future, we will therefore examine the suitability of our developed embedded expression language with respect to the mapping of relevant concepts from a roadmap engineer's point of view.
To enable this, we are currently working on a modeling methodology for \TextIris\ that supports domain experts in deciding how to use the presented domain-specific modeling language and our tool. Project partners have recently published a modeling methodology in~\cite{Fakih2021} that is independent from the specifics of our domain-specific language. We are currently collaborating with them to identify how well their modeling methodology aligns with our language. A particularly interesting aspect is how detailed each company in a value chain models its relevant aspects as well as when and how often models are exchanged between  partners in the value chain.

Finally, we will take a closer look at the visualization and exploration of the model-wide solution space.

\section*{Acknowledgements}

This work has been developed in the project GENIAL! (reference number: 16ES0875). GENIAL! is partly funded by the German Federal Ministry of Education and Research (BMBF) within the research programme ICT 2020.

Further, this work was partially funded by the Deutsche Forschungsgemeinschaft (DFG, German Research Foundation) -- 453895475.

We would like to thank our academic and industrial project partners for extensive discussions and valuable insights.

In the course of this article, several screenshots of \TextIris\ are shown. The screenshots include icons of the following authors:

\begin{itemize}
    \item Some icons by Yusuke Kamiyamane\\*
    \url{https://p.yusukekamiyamane.com/}\\
    Licensed under a Creative Commons Attribution 3.0 License\\ (\url{https://creativecommons.org/licenses/by/3.0/}).
    \item Some icons are taken from the Silk Icon Set\\*
    \url{http://www.famfamfam.com/lab/icons/silk/}\\
    Licensed under a Creative Commons Attribution 3.0 License (see URL above).
    \item Further, some icons are taken from FatCow\\*
    \url{https://www.fatcow.com/free-icons}\\
    Licensed under a Creative Commons Attribution 3.0 United States License\\ \url{https://creativecommons.org/licenses/by/3.0/us/}
\end{itemize}

\appendix
\section{Generated Constraint System for Figure~\ref{fig:runingExample}\label{appendix:constraint-system}}
{\footnotesize
\begin{iris}[numbers=left,stepnumber=5]
AutonomousDriving.?requirement1(T) = PowerSupply.?availability(T)
AutonomousDriving.?requirement2(T) = ErrorDetection.?availability(T)
AutonomousDriving.?availability(T) = AutonomousDriving.?requirement1(T) & AutonomousDriving.?requirement2(T)
AutonomousDriving.?replacement(T) = -1
Vehicle.TotalCurrent(T) = Headlights.Current(T) + ProcessingUnits.Current(T)
Vehicle.?availability(T) = Headlights.?availability(T) & DetectionSoftware.?availability(T) & Autopilot.?availability(T) & ProcessingUnits.?availability(T) & Fuse.?availability(T)
Vehicle.?replacement(T) = -1
Headlights.Current(T) = 5A
Headlights.?availability(T) = true
Headlights.?replacement(T) = -1
DetectionSoftware.TFLOPS(T) = linear(T, Jan2021, 100, Jan2035, 200)
DetectionSoftware.?availability(T) = true
DetectionSoftware.?replacement(T) = -1
Autopilot.TFLOPS(T) = 50
Autopilot.?availability(T) = true
Autopilot.?replacement(T) = -1
ProcessingUnits.PowerConsumption(T) = ((DetectionSoftware.TFLOPS(T) + Autopilot.TFLOPS(T))/80) * 200W
ProcessingUnits.Current(T) = (ProcessingUnits.PowerConsumption(T) / 12V)[A]
ProcessingUnits.?availability(T) = true
ProcessingUnits.?replacement(T) = -1
Fuse.?kpi1(BlFuse,T) = num(BlFuse.Watchdog(T))
Fuse.?kpi1(EFuse,T) = num(EFuse.Watchdog(T))
Fuse.MaxLoadCurrent(T) =
         if Fuse.?replacement(T) = 1 then BlFuse.MaxLoadCurrent(T)
    else if Fuse.?replacement(T) = 2 then EFuse.MaxLoadCurrent(T)
    else [-inf..inf]*1A
Fuse.Watchdog(T) =
         if Fuse.?replacement(T) = 1 then BlFuse.Watchdog(T)
    else if Fuse.?replacement(T) = 2 then EFuse.Watchdog(T)
    else maybe
Fuse.?requirement1(T) = (Fuse.MaxLoadCurrent(T) >= Vehicle.TotalCurrent(T))
Fuse.BatteryVoltage(T) =
         if Fuse.?replacement(T) = 1 then BlFuse.BatteryVoltage(T)
    else if Fuse.?replacement(T) = 2 then EFuse.BatteryVoltage(T)
    else 48V
Fuse.?availability(T) = Fuse.?requirement1(T)
       & if Fuse.?replacement(T) = 1 then BlFuse.?availability(T)
    else if Fuse.?replacement(T) = 2 then EFuse.?availability(T)
    else true
Fuse.?replacement(T) = index_of_max(
   if BlFuse.?availability(T) then Fuse.?kpi1(BlFuse,T) else -inf,
   if EFuse.?availability(T) then Fuse.?kpi1(EFuse,T) else -inf
)
PowerSupply.?requirement1(T) = (Fuse.MaxLoadCurrent(T) >= Vehicle.TotalCurrent(T))
PowerSupply.?availability(T) = PowerSupply.?requirement1
PowerSupply.?replacement(T) = -1
ErrorDetection.?requirement1(T) = Fuse.Watchdog(T)
ErrorDetection.?availability(T) = ErrorDetection.?requirement1(T)
ErrorDetection.?replacement(T) = -1
BlFuse.?requirement1(T) = (BlFuse.MaxLoadCurrent(T) >= Vehicle.TotalCurrent(T))
BlFuse.BatteryVoltage(T) = 48V
BlFuse.MaxLoadCurrent(T) = 50A
BlFuse.Watchdog(T) = false
BlFuse.?availability(T) = BlFuse.?requirement1(T)
BlFuse.?replacement(T) = -1
EFuse.?requirement1(T) = (EFuse.MaxLoadCurrent(T) >= Vehicle.TotalCurrent(T))
EFuse.BatteryVoltage(T) = 48V
EFuse.MaxLoadCurrent(T) = 45A
EFuse.Watchdog(T) = true
EFuse.?requirement2(T) = (T >= Jan2022 + [months(12) .. months(36)])
EFuse.?availability(T) = EFuse.?requirement1(T) & EFuse.?requirement2(T)
EFuse.?replacement(T) = -1
\end{iris}
}

\section{Symbolic Transformations\label{sec:symbolic-trans}}

{%
\newcommand*{\irisexample}[2]{\item #1 $\Longrightarrow$ #2}%
\newlist{irisexamples}{itemize}{1}
\setlist[irisexamples,1]{label={--},leftmargin=*,before={\leavevmode\\*\textit{Example:}\nobreak},topsep=0pt,partopsep=0pt}
\begin{itemize}
    \item \textbf{Constant Folding}. Subexpressions involving only constant values are reduced by evaluating them as far as possible.
    \begin{irisexamples}
        \irisexample{\irisinline{if max(2, 3) >= 2.5 then 1 else 2}}{\irisinline{1}}
    \end{irisexamples}
    \item \textbf{Propagation}. References are replaced by their inferred values.
    \begin{irisexamples}
        \irisexample{\irisinline{x = [1..5] & y = x}}{\irisinline{x = [1..5] & y = x & y = [1..5]}}
    \end{irisexamples}
    Note that in case of uncertain intervals the original subexpression \irisinline{y = x} remains in case stronger bounds for \texttt{x} are inferred later on.

    \item \textbf{Neutral Element Removal}. Neutral parts of expressions are removed.
    \begin{irisexamples}
        \irisexample{\irisinline{x = (y & true)}}{\irisinline{x = y}}
    \end{irisexamples}
    \item \textbf{Reordering}. Commutative operations are reordered to achieve a normal form where possible and allow for subsequent simplifications.
    \begin{irisexamples}
        \irisexample{\irisinline{x + 2 = -x + 3}}{\irisinline{x + x = 3 - 2}}
    \end{irisexamples}

    \item \textbf{Raising/Lowering}. Subexpressions are propagated up and down the syntax tree to allow for subsequent simplifications.
    \begin{irisexamples}
        \irisexample{\irisinline{2 * (if x then 3 else 4)}}{\irisinline{if x then 2 * 3 else 2 * 4}}
    \end{irisexamples}

    \item \textbf{Merging}. Identical subexpressions are merged where possible.
    \begin{irisexamples}
        \irisexample{\irisinline{y + y}}{\irisinline{2 * y}}
        \irisexample{\irisinline{if z then x else x}}{\irisinline{x}}
        \irisexample{\irisinline{(x < y) & (if x < y then z1 else z2)}}{\irisinline{(x < y) & z1}}
    \end{irisexamples}

    \item \textbf{Special Case Detection}. Various properties of functions are used to simplify commonly occurring cases.
    \begin{irisexamples}
        \irisexample{\irisinline{linear(x, x1, 0, x2, 1) >= 0}}{\irisinline{true}}
     \end{irisexamples}
\end{itemize}
}

\section{More Details on the Domain-Specific Language}

\subsection{General Syntactic Elements of the Expression Language}

Our textual expression language provides the following general language elements. Sections~\ref{sec:time-dependent-properties} and~\ref{sec:uncertainty} extend this set of elements with dates, durations, and uncertainty.

\label{app:syntax}\begin{itemize}
    \item Boolean constants, i.e., \texttt{true} and \texttt{false}
    \item Numeric constants, optionally suffixed by a SI-unit, e.g., \irisinline{70.5}, \irisinline{12V}, \irisinline{400mA}
    \item The constant \irisinline{inf}, denoting positive infinity
    \item Interval expressions, i.e., \texttt{[\textit{lower}..\textit{upper}]}, specifying a closed interval of values that will be propagated through all arithmetic operations. We use intervals to represent uncertain values that are known to not exceed a certain interval, e.g., \irisinline{[1.5..2.5]}. Section~\ref{sec:uncertainty} contains a more detailed description of arithmetic operations performed on intervals.
    \item Identifiers which reference the properties and blocks specified in the model, e.g., \texttt{TotalCurrent}, or \texttt{Vehicle.Fuse.Watchdog}. If an identifier references a block, the resulting value is the boolean availability of the block. If an identifier references a property, the resulting value is the solver result of the property formula.
    \item Arithmetic operators \{\texttt{+}, \texttt{-}, \texttt{*}, \texttt{/}, \irisinline{^}\}, e.g., \irisinline{2 * 12V}
    \item Relational operators \{\texttt{<}, \texttt{<=}, \texttt{>}, \texttt{>=}, \texttt{==}, \texttt{!=}\}, e.g., \irisinline{x > y}
    \item Boolean operators \{\texttt{\&}, \texttt{|}, \texttt{!}\}, e.g., \irisinline{a & !b}
    \item Parenthesis, e.g., \irisinline{a * (b + c)}.
    \item Conditional expressions \irisinline{if} \textit{\texttt{c}} \irisinline{then} \textit{\texttt{t}} \irisinline{else} \textit{\texttt{e}}, which evaluate to \textit{\texttt{t}} if the condition \textit{\texttt{c}} evaluates to $true$, and \textit{\texttt{e}} otherwise.
    \item Aggregations \{\irisinline{SUM}, \irisinline{PRODUCT}, \irisinline{AND}, \irisinline{OR}, \irisinline{MIN}, \irisinline{MAX}, \irisinline{UNION}\}. An aggregation, like \texttt{SUM(\textit{expr})}, evaluates \texttt{\textit{expr}} for all direct children of the block containing the expression, and computes a function, like $\Sigma$, over their values. Aggregations are used in roadmap models to construct parent blocks that combine properties of their children in a uniform way, like computing the total current of all subsystems.
    \item Utility functions \{\irisinline{sin}, \irisinline{cos}, \irisinline{exp}, \irisinline{ln}, \irisinline{log}, \irisinline{sqrt}, \irisinline{min}, \irisinline{max}, \irisinline{num}\}. The syntax for function calls is \irisinline{f(arg1, ..., argn)}. The set of functions was defined based on needs during our tests and evaluations, and can be extended easily.
    \item An interpolation function \texttt{linear(\textit{x}, $\textit{\texttt{x}}_\textit{0}$, $\textit{\texttt{y}}_\textit{0}$, ..., $\textit{\texttt{x}}_\textit{n}$, $\textit{\texttt{y}}_\textit{n}$)}, which computes the linearly interpolated value \textit{\texttt{y}} at position \textit{\texttt{x}} in the sequence of \texttt{($\textit{\texttt{x}}_\textit{i}$,$\textit{\texttt{y}}_\textit{i}$)}-values. The \textit{\texttt{x}}-values are usually instantiated with time-based values, which will be introduced in the next section. This \texttt{linear}-function in particular could also be defined using nested conditional expressions, but we include it as a concise and intuitive way of specifying values that vary over time.
\end{itemize}

\subsection{Arithmetic and Relational Operations on Dates and Durations}\label{app:date-ops}
In our language, we use the following arithmetic and relational operations and typing rules ($\circ$ represents the relational operators $<$, $\leq$, $>$, $\geq$, $=$, $\neq$):
\newcommand\timesdiv{\mathbin{\vcenter{\hbox{%
   $\begin{array}{@{}c@{}}\times\\[-2.667ex]\div\end{array}$}}}}
 \newcommand*\tirisinline[1]{\text{\irisinline{#1}}}
\begin{alignat*}{3}
    date     &\pm    duration &&:= date     & (\tirisinline{Jan2021 + months(5) = Jun2021}) \\
    duration &+      date     &&:= date     & (\tirisinline{months(5) + Jan2021 = Jun2021}) \\
    date     &-      date     &&:= duration & (\tirisinline{Jun2021 - Jan2021 = months(5)}) \\
    duration &\pm    duration &&:= duration & ~~(\tirisinline{months(2) + months(3) = months(5)}) \\
    duration &\times number   &&:= duration & (\tirisinline{months(2) * 3 = months(6)}) \\
    number   &\times duration &&:= duration & (\tirisinline{3 * months(2) = months(6)}) \\
    duration &\div   number   &&:= duration & (\tirisinline{months(6) / 3 = months(2)}) \\
    date     &\circ  date     &&:= boolean & (\tirisinline{Jan2021 <= Jun2021 = true}) \\
    duration &\circ  duration &&:= boolean & (\tirisinline{months(4) <= months(2) = false})
\end{alignat*}

Note that $date - date$ computes the duration between the two dates, whereas other combinations, like $date + date$, are explicitly disallowed as they would also depend on the neutral element of the underlying calendar system (i.e., year 0). $duration + date$ however is allowed to ensure commutativity of the $+$ operator.

\subsection{More on Interval Operations}
\label{app:interval-ops}This subsection provides a more detailed look at the behavior of interval operations in the domain-specific language. The behavior should be intuitive to those who are already familiar with interval operations.\par

\let\varstyle\texttt\def\a{\varstyle{a}}\def\b{\varstyle{b}}\def\c{\varstyle{c}}\def\d{\varstyle{d}}%
\def\T#1#2{\texttt{[\ensuremath{#1}..\ensuremath{#2}]}}%
\def\RT#1{\texttt{[\ensuremath{#1}..\(\infty\))}}%
\def\LT#1{\texttt{(\(-\infty\)..\ensuremath{#1}]}}%
\def\InfT{\texttt{(\(-\infty\)..\(\infty\))}}%
Let \a, \b, \c\ and \d\ be numeric constants, as defined in Section~\ref{sec:exprlng}, \(\circ\) represents any of the basic arithmetic operators: \(+\), \(-\), \(*\), and \(\div\):
\begin{alignat*}{2}
    \T{\a}{\b}     &\circ     \c &&= \T{\a\circ\c}{\b\circ\c}\\
    \T{\a}{\b}     &+ \T{\c}{\d} &&= \T{\a+\c}{\b+\d}\\
    \T{\a}{\b}     &- \T{\c}{\d} &&= \T{\a-\d}{\b-\c}\\
    \T{\a}{\b}     &* \T{\c}{\d} &&= \T{\min\{\a\c,\a\d,\b\c,\b\d\}}{\max\{\a\c,\a\d,\b\c,\b\d\}}\\
    \T{\a}{\b}     &/ \T{\c}{\d} &&= \T{\a}{\b} * \frac{1}{\T{\c}{\d}} \\
    1              &/ \T{\a}{\b} &&= \begin{cases}
                                        \LT{1/\a}, & \text{if } \b = 0 \\
                                        \RT{1/\b}, & \text{if } \a = 0 \\
                                        \LT{1/\a} \cup \RT{1/\b}, & \text{if } \a < 0 < \b \\
                                        \T{1/\a}{1/\b}, & \text{otherwise}
                                     \end{cases}
\end{alignat*}
After each operation the invariant \(\a \leq \b\) is ensured by swapping the resulting bounds \a\ and \b\ if necessary.
The split interval \(\LT{1/\c} \cup \RT{1/\d}\) for \(\c < 0 < \d\) will be joined in order to create a single consecutive interval: \(\T{\a}{\b} \cup \T{\c}{\d} = \T{\a}{\d}\). Therefore a division like \texttt{\T{1}{2}/\T{-1}{1}} yields the result \InfT, because \(0 \in \T{-1}{1}\) leads to the following multiplication
\begin{align*}
\T{1}{2} * \InfT &= [\, \min\{1 * (-\infty),\;1 * \infty,\;2 * (-\infty),\;2 * \infty\} \\
                &\qquad \texttt{..}\, \max\{1 * (-\infty),\;1 * \infty,\;2 * (-\infty),\;2 * \infty\}\,] \\
  &= \InfT    
\end{align*}

\bibliography{literature}

\begin{thebibliography}{10}
\expandafter\ifx\csname url\endcsname\relax
  \def\url#1{\texttt{#1}}\fi
\expandafter\ifx\csname urlprefix\endcsname\relax\def\urlprefix{URL }\fi
\expandafter\ifx\csname href\endcsname\relax
  \def\href#1#2{#2} \def\path#1{#1}\fi

\bibitem{EdisonAT13}
H.~Edison, N.~B. Ali, R.~Torkar, Towards innovation measurement in the software
  industry, Journal of Systems and Software 86~(5) (2013) 1390--1407.
\newblock \href {http://dx.doi.org/10.1016/j.jss.2013.01.013}
  {\path{doi:10.1016/j.jss.2013.01.013}}.

\bibitem{Satell2017}
G.~Satell,
  \href{https://hbr.org/2017/06/the-4-types-of-innovation-and-the-problems-they-solve}{The
  4 types of innovation and the problems they solve} (2017).
\newline\urlprefix\url{https://hbr.org/2017/06/the-4-types-of-innovation-and-the-problems-they-solve}

\bibitem{Kerr.2020}
C.~Kerr, R.~Phaal, Technology roadmapping: Industrial roots, forgotten history
  and unknown origins, Technological Forecasting and Social Change 155 (2020)
  1--16.
\newblock \href
  {http://dx.doi.org/https://doi.org/10.1016/j.techfore.2020.119967}
  {\path{doi:https://doi.org/10.1016/j.techfore.2020.119967}}.

\bibitem{Park.2020}
H.~Park, R.~Phaal, J.-Y. Ho, E.~O'Sullivan,
  \href{https://www.sciencedirect.com/science/article/pii/S0040162519304901}{Twenty
  years of technology and strategic roadmapping research: A school of thought
  perspective}, Technological Forecasting and Social Change 154 (2020) 119965.
\newblock \href
  {http://dx.doi.org/https://doi.org/10.1016/j.techfore.2020.119965}
  {\path{doi:https://doi.org/10.1016/j.techfore.2020.119965}}.
\newline\urlprefix\url{https://www.sciencedirect.com/science/article/pii/S0040162519304901}

\bibitem{Willyard.1987}
C.~H. Willyard, C.~W. McClees, Motorola's technology roadmap process, Research
  Management 30~(5) (1987) 13--19.
\newblock \href {http://dx.doi.org/10.1080/00345334.1987.11757057}
  {\path{doi:10.1080/00345334.1987.11757057}}.

\bibitem{Letaba.2015}
P.~Letaba, M.~W. Pretorius, L.~Pretorius, Analysis of the intellectual
  structure and evolution of technology roadmapping literature, in: 2015
  Portland International Conference on Management of Engineering and Technology
  (PICMET), 2015, pp. 2248--2254.
\newblock \href {http://dx.doi.org/10.1109/PICMET.2015.7273147}
  {\path{doi:10.1109/PICMET.2015.7273147}}.

\bibitem{Vatananan.2012}
R.~S. Vatananan, N.~Gerdsri, {The Current State of Technology Roadmapping (TRM)
  Research and Practice}, {International Journal of Innovation Technol.
  Management} 9~(4) (2012) 1--20.
\newblock \href {http://dx.doi.org/10.1142/S0219877012500320}
  {\path{doi:10.1142/S0219877012500320}}.

\bibitem{Rinne.2004}
M.~Rinne, {Technology roadmaps: Infrastructure for innovation}, {Technological
  Forecasting and Social Change} 71~(1-2) (2004) 67--80.
\newblock \href {http://dx.doi.org/10.1016/j.techfore.2003.10.002}
  {\path{doi:10.1016/j.techfore.2003.10.002}}.

\bibitem{Garcia.1997}
M.~L. Garcia, Introduction to Technology Roadmapping: The Semiconductor
  Industry Association’s Technology Roadmapping Process, Sandia National
  Laboratories, 1997.

\bibitem{Holmes.2006}
C.~Holmes, M.~{A Ferrill}, {A Process for the Update and Review of Operation
  and Technology Roadmaps}, in: {Proceedings of the IEEE International
  Conference on Management of Innovation and Technology (ICMIT'06)}, {IEEE
  Operations Center}, 2006, pp. 984--988.
\newblock \href {http://dx.doi.org/10.1109/ICMIT.2006.262369}
  {\path{doi:10.1109/ICMIT.2006.262369}}.

\bibitem{Lee.2009}
J.~Lee, C.-Y. Lee, T.-Y. Kim, A practical approach for beginning the process of
  technology roadmapping, International Journal of Technology Management 47~(4)
  (2009) 306--321.

\bibitem{Martin.2012}
H.~Martin, T.~U. Daim, Technology roadmap development process (trdp) for the
  service sector: A conceptual framework, Technology in Society 34~(1) (2012)
  94--105.

\bibitem{Deursen.2000}
A.~{van Deursen}, P.~Klint, J.~Visser, {Domain-Specific Languages: An Annotated
  Bibliography}, {ACM Sigplan Notices} 35~(6) (2000) 26--36.

\bibitem{Fowler.2011}
M.~Fowler, Domain-Specific Languages, Addison-Wesley, 2011.

\bibitem{OMGSysML15}
{Object Management Group (OMG)}, \href{http://www.omg.org/spec/SysML/1.5/}{{OMG
  Systems Modeling Language (OMG SysML), Version 1.5}}, Standard, OMG (2017).
\newline\urlprefix\url{http://www.omg.org/spec/SysML/1.5/}

\bibitem{SEAA2020}
A.~Breckel, J.~Pietron, K.~Juhnke, M.~Tichy, A domain-specific language and
  interactive user interface for model-driven engineering of technology
  roadmaps, in: Proceedings of the 46th Euromicro Conference on Software
  Engineering and Advanced Applications (SEAA'20), 2020, pp. 162--170.
\newblock \href {http://dx.doi.org/10.1109/SEAA51224.2020.00035}
  {\path{doi:10.1109/SEAA51224.2020.00035}}.

\bibitem{Phaal.2004}
R.~Phaal, C.~J. Farrukh, D.~R. Probert, {Technology roadmapping -- A planning
  framework for evolution and revolution}, {Technological Forecasting and
  Social Change} 71~(1-2) (2004) 5--26.
\newblock \href {http://dx.doi.org/10.1016/S0040-1625(03)00072-6}
  {\path{doi:10.1016/S0040-1625(03)00072-6}}.

\bibitem{Kostoff.2001}
R.~N. Kostoff, R.~R. Schaller, {Science and Technology Roadmaps}, {IEEE
  Transactions on Engineering Management} 48~(2) (2001) 132--143.
\newblock \href {http://dx.doi.org/10.1109/17.922473}
  {\path{doi:10.1109/17.922473}}.

\bibitem{Knoll.2018}
D.~Knoll, A.~Golkar, O.~de~Weck, {A concurrent design approach for model-based
  technology roadmapping}, in: {The 12th Annual IEEE International Systems
  Conference (SysCon'18)}, IEEE, Piscataway, NJ, USA, 2018, pp. 1--6.
\newblock \href {http://dx.doi.org/10.1109/SYSCON.2018.8369527}
  {\path{doi:10.1109/SYSCON.2018.8369527}}.

\bibitem{ISO.19450}
ISO, \href{https://www.iso.org/standard/62274.html}{Automation systems and
  integration -- object-process methodology}, ISO/PAS 19450:2015, ISO (2015).
\newline\urlprefix\url{https://www.iso.org/standard/62274.html}

\bibitem{CR05}
F.~Cassez, O.-H. Roux, Structural translation from time petri nets to timed
  automata, Electronic Notes in Theoretical Computer Science 128 (2005)
  145--160.

\bibitem{Alur&Dill1994}
R.~Alur, D.~L. Dill, {A} {T}heory of {T}imed {A}utomata, Theoretical Computer
  Science 126 (1994) 183--235.

\bibitem{OMG2009}
{Object Management Group (OMG)}, \href{https://www.omg.org/spec/MARTE/}{{UML
  Profile for MARTE™: Modeling and Analysis of Real-Time Embedded Systems
  Version 1.2}}, Standard, OMG (2019).
\newline\urlprefix\url{https://www.omg.org/spec/MARTE/}

\bibitem{lee2010bayesian}
C.~Lee, B.~Song, Y.~Cho, Y.~Park, A bayesian belief network approach to
  operationalization of multi-scenario technology roadmap, in: PICMET 2010
  TECHNOLOGY MANAGEMENT FOR GLOBAL ECONOMIC GROWTH, IEEE, 2010, pp. 1--6.

\bibitem{jeong2021developing}
Y.~Jeong, H.~Jang, B.~Yoon, Developing a risk-adaptive technology roadmap using
  a bayesian network and topic modeling under deep uncertainty, Scientometrics
  126~(5) (2021) 3697--3722.

\bibitem{kleene1938notation}
S.~C. Kleene, On notation for ordinal numbers, The Journal of Symbolic Logic
  3~(4) (1938) 150--155.

\bibitem{moore1966interval}
R.~E. Moore, Interval analysis, Vol.~4, Prentice-Hall Englewood Cliffs, 1966.

\bibitem{DBLP:journals/tse/MedvidovicT00}
N.~Medvidovic, R.~N. Taylor, \href{https://doi.org/10.1109/32.825767}{A
  classification and comparison framework for software architecture description
  languages}, {IEEE} Trans. Software Eng. 26~(1) (2000) 70--93.
\newblock \href {http://dx.doi.org/10.1109/32.825767}
  {\path{doi:10.1109/32.825767}}.
\newline\urlprefix\url{https://doi.org/10.1109/32.825767}

\bibitem{DBLP:journals/sttt/MkaouarZHJ20}
H.~Mkaouar, B.~Zalila, J.~Hugues, M.~Jmaiel,
  \href{https://doi.org/10.1007/s10009-019-00513-7}{A formal approach to {AADL}
  model-based software engineering}, Int. J. Softw. Tools Technol. Transf.
  22~(2) (2020) 219--247.
\newblock \href {http://dx.doi.org/10.1007/s10009-019-00513-7}
  {\path{doi:10.1007/s10009-019-00513-7}}.
\newline\urlprefix\url{https://doi.org/10.1007/s10009-019-00513-7}

\bibitem{Bao+2017}
Y.~Bao, M.~Chen, Q.~Zhu, T.~Wei, F.~Mallet, T.~Zhou, Quantitative performance
  evaluation of uncertainty-aware hybrid aadl designs using statistical model
  checking, IEEE Transactions on Computer-Aided Design of Integrated Circuits
  and Systems 36~(12) (2017) 1989--2002.
\newblock \href {http://dx.doi.org/10.1109/TCAD.2017.2681076}
  {\path{doi:10.1109/TCAD.2017.2681076}}.

\bibitem{DBLP:journals/tse/AletiBGKM13}
A.~Aleti, B.~Buhnova, L.~Grunske, A.~Koziolek, I.~Meedeniya,
  \href{https://doi.org/10.1109/TSE.2012.64}{Software architecture optimization
  methods: {A} systematic literature review}, {IEEE} Trans. Software Eng.
  39~(5) (2013) 658--683.
\newblock \href {http://dx.doi.org/10.1109/TSE.2012.64}
  {\path{doi:10.1109/TSE.2012.64}}.
\newline\urlprefix\url{https://doi.org/10.1109/TSE.2012.64}

\bibitem{DBLP:books/daglib/0032924}
S.~Apel, D.~S. Batory, C.~K{\"{a}}stner, G.~Saake,
  \href{https://doi.org/10.1007/978-3-642-37521-7}{Feature-Oriented Software
  Product Lines - Concepts and Implementation}, Springer, 2013.
\newblock \href {http://dx.doi.org/10.1007/978-3-642-37521-7}
  {\path{doi:10.1007/978-3-642-37521-7}}.
\newline\urlprefix\url{https://doi.org/10.1007/978-3-642-37521-7}

\bibitem{DBLP:journals/tosem/HieronsLLSZ16}
R.~M. Hierons, M.~Li, X.~Liu, S.~Segura, W.~Zheng,
  \href{https://doi.org/10.1145/2897760}{{SIP:} optimal product selection from
  feature models using many-objective evolutionary optimization}, {ACM} Trans.
  Softw. Eng. Methodol. 25~(2) (2016) 17:1--17:39.
\newblock \href {http://dx.doi.org/10.1145/2897760}
  {\path{doi:10.1145/2897760}}.
\newline\urlprefix\url{https://doi.org/10.1145/2897760}

\bibitem{graf_sense_1996}
A.~Graf, A.~Koroncal, P.~Sommer, J.~Tihanyi, Sense highside switch in smart
  power technology takes over fuse function, in: 7. Internationale Fachtagung
  Elektronik im Kraftfahrzeug, 1996, pp. 331--344.

\bibitem{fuisting_current_2014}
M.~Fuisting, R.~Gillhaus, J.~Olk, T.~Gauter, H.~Völkel, J.~Falkenstein,
  Current {Switch} and {Sense} {Module} for the intelligent power distribution
  in future {E}-/{E}-architectures, in: Automotive meets Electronics (AmE'14),
  VDE Verlag, 2014, pp. 1--6.

\bibitem{10.1145/1995376.1995394}
L.~De~Moura, N.~Bj\o{}rner,
  \href{https://doi.org/10.1145/1995376.1995394}{Satisfiability modulo
  theories: Introduction and applications}, Commun. ACM 54~(9) (2011) 69–77.
\newblock \href {http://dx.doi.org/10.1145/1995376.1995394}
  {\path{doi:10.1145/1995376.1995394}}.
\newline\urlprefix\url{https://doi.org/10.1145/1995376.1995394}

\bibitem{OCL2014}
{Object Management Group (OMG)}, \href{http://www.omg.org/spec/OCL/2.4}{{Object
  Constraint Language (OCL) Specification Version 2.4}}, Standard, OMG (2014).
\newline\urlprefix\url{http://www.omg.org/spec/OCL/2.4}

\bibitem{10.1145/3377024.3377036}
J.~Sprey, C.~Sundermann, S.~Krieter, M.~Nieke, J.~Mauro, T.~Th\"{u}m,
  I.~Schaefer, \href{https://doi.org/10.1145/3377024.3377036}{Smt-based
  variability analyses in featureide}, in: Proceedings of the 14th
  International Working Conference on Variability Modelling of
  Software-Intensive Systems, VAMOS '20, Association for Computing Machinery,
  New York, NY, USA, 2020, pp. 1--9.
\newblock \href {http://dx.doi.org/10.1145/3377024.3377036}
  {\path{doi:10.1145/3377024.3377036}}.
\newline\urlprefix\url{https://doi.org/10.1145/3377024.3377036}

\bibitem{10.1016/j.is.2010.01.001}
D.~Benavides, S.~Segura, A.~Ruiz-Cort\'{e}s,
  \href{https://doi.org/10.1016/j.is.2010.01.001}{Automated analysis of feature
  models 20 years later: A literature review}, Inf. Syst. 35~(6) (2010)
  615–636.
\newblock \href {http://dx.doi.org/10.1016/j.is.2010.01.001}
  {\path{doi:10.1016/j.is.2010.01.001}}.
\newline\urlprefix\url{https://doi.org/10.1016/j.is.2010.01.001}

\bibitem{neumaier_1991}
A.~Neumaier, Interval Methods for Systems of Equations, Encyclopedia of
  Mathematics and its Applications, Cambridge University Press, 1991.
\newblock \href {http://dx.doi.org/10.1017/CBO9780511526473}
  {\path{doi:10.1017/CBO9780511526473}}.

\bibitem{Sprey2018ComputingAR}
J.~Sprey, C.~Sundermann, Computing attribute ranges for partial configurations
  with javasmt: Bachelor's thesis, Master's thesis, Technische Universität
  Braunschweig (2018).

\bibitem{DBLP:conf/tacas/MouraB08}
L.~M. de~Moura, N.~Bj{\o}rner,
  \href{https://doi.org/10.1007/978-3-540-78800-3\_24}{{Z3:} an efficient {SMT}
  solver}, in: C.~R. Ramakrishnan, J.~Rehof (Eds.), Tools and Algorithms for
  the Construction and Analysis of Systems, 14th International Conference,
  {TACAS} 2008, Held as Part of the Joint European Conferences on Theory and
  Practice of Software, {ETAPS} 2008, Budapest, Hungary, March 29-April 6,
  2008. Proceedings, Vol. 4963 of Lecture Notes in Computer Science, Springer,
  2008, pp. 337--340.
\newblock \href {http://dx.doi.org/10.1007/978-3-540-78800-3\_24}
  {\path{doi:10.1007/978-3-540-78800-3\_24}}.
\newline\urlprefix\url{https://doi.org/10.1007/978-3-540-78800-3\_24}

\bibitem{DBLP:conf/lpar/MouraB08}
L.~M. de~Moura, N.~Bj{\o}rner,
  \href{http://ceur-ws.org/Vol-418/paper10.pdf}{Proofs and refutations, and
  {Z3}}, in: P.~Rudnicki, G.~Sutcliffe, B.~Konev, R.~A. Schmidt, S.~Schulz
  (Eds.), Proceedings of the {LPAR} 2008 Workshops, Knowledge Exchange:
  Automated Provers and Proof Assistants, and the 7th International Workshop on
  the Implementation of Logics, Doha, Qatar, November 22, 2008, Vol. 418 of
  {CEUR} Workshop Proceedings, CEUR-WS.org, 2008, pp. 123--132.
\newline\urlprefix\url{http://ceur-ws.org/Vol-418/paper10.pdf}

\bibitem{DBLP:books/sp/NipkowPW02}
T.~Nipkow, L.~C. Paulson, M.~Wenzel,
  \href{https://doi.org/10.1007/3-540-45949-9}{Isabelle/HOL - {A} Proof
  Assistant for Higher-Order Logic}, Vol. 2283 of Lecture Notes in Computer
  Science, Springer, 2002.
\newblock \href {http://dx.doi.org/10.1007/3-540-45949-9}
  {\path{doi:10.1007/3-540-45949-9}}.
\newline\urlprefix\url{https://doi.org/10.1007/3-540-45949-9}

\bibitem{DBLP:journals/jpl/Tip95}
F.~Tip, \href{http://compscinet.dcs.kcl.ac.uk/JP/jp030301.abs.html}{A survey of
  program slicing techniques}, J. Program. Lang. 3~(3).
\newline\urlprefix\url{http://compscinet.dcs.kcl.ac.uk/JP/jp030301.abs.html}

\bibitem{DBLP:journals/sttt/BurmesterGNTWWWZ04}
S.~Burmester, H.~Giese, J.~Niere, M.~Tichy, J.~P. Wadsack, R.~Wagner,
  L.~Wendehals, A.~Z{\"{u}}ndorf,
  \href{https://doi.org/10.1007/s10009-004-0155-8}{Tool integration at the
  meta-model level: the fujaba approach}, Int. J. Softw. Tools Technol. Transf.
  6~(3) (2004) 203--218.
\newblock \href {http://dx.doi.org/10.1007/s10009-004-0155-8}
  {\path{doi:10.1007/s10009-004-0155-8}}.
\newline\urlprefix\url{https://doi.org/10.1007/s10009-004-0155-8}

\bibitem{DBLP:conf/dagstuhl/PriesterjahnTHHS07}
C.~Priesterjahn, M.~Tichy, S.~Henkler, M.~Hirsch, W.~Sch{\"{a}}fer,
  \href{https://doi.org/10.1007/978-3-642-16277-0\_12}{Fujaba4eclipse real-time
  tool suite}, in: H.~Giese, G.~Karsai, E.~Lee, B.~Rumpe, B.~Sch{\"{a}}tz
  (Eds.), Model-Based Engineering of Embedded Real-Time Systems - International
  Dagstuhl Workshop, Dagstuhl Castle, Germany, November 4-9, 2007. Revised
  Selected Papers, Vol. 6100 of Lecture Notes in Computer Science, Springer,
  2007, pp. 309--315.
\newblock \href {http://dx.doi.org/10.1007/978-3-642-16277-0\_12}
  {\path{doi:10.1007/978-3-642-16277-0\_12}}.
\newline\urlprefix\url{https://doi.org/10.1007/978-3-642-16277-0\_12}

\bibitem{DBLP:conf/sle/MaroSATG15}
S.~Maro, J.~Stegh{\"{o}}fer, A.~Anjorin, M.~Tichy, L.~Gelin,
  \href{https://dl.acm.org/citation.cfm?id=2814253}{On integrating graphical
  and textual editors for a {UML} profile based domain specific language: an
  industrial experience}, in: R.~F. Paige, D.~D. Ruscio, M.~V{\"{o}}lter
  (Eds.), Proceedings of the 2015 {ACM} {SIGPLAN} International Conference on
  Software Language Engineering, {SLE} 2015, Pittsburgh, PA, USA, October
  25-27, 2015, {ACM}, 2015, pp. 1--12.
\newline\urlprefix\url{https://dl.acm.org/citation.cfm?id=2814253}

\bibitem{DBLP:conf/gg/0001BGGKOT17}
D.~Str{\"{u}}ber, K.~Born, K.~D. Gill, R.~Groner, T.~Kehrer, M.~Ohrndorf,
  M.~Tichy, \href{https://doi.org/10.1007/978-3-319-61470-0\_12}{Henshin: {A}
  usability-focused framework for {EMF} model transformation development}, in:
  J.~de~Lara, D.~Plump (Eds.), Graph Transformation - 10th International
  Conference, {ICGT} 2017, Held as Part of {STAF} 2017, Marburg, Germany, July
  18-19, 2017, Proceedings, Vol. 10373 of Lecture Notes in Computer Science,
  Springer, 2017, pp. 196--208.
\newblock \href {http://dx.doi.org/10.1007/978-3-319-61470-0\_12}
  {\path{doi:10.1007/978-3-319-61470-0\_12}}.
\newline\urlprefix\url{https://doi.org/10.1007/978-3-319-61470-0\_12}

\bibitem{DBLP:conf/staf/TichyMJH20}
M.~Tichy, J.~Pietron, D.~M{\"{o}}dinger, K.~Juhnke, F.~J. Hauck,
  \href{http://ceur-ws.org/Vol-2707/mde4iotpaper1.pdf}{Experiences with an
  internal {DSL} in the iot domain}, in: L.~Burgue{\~{n}}o, L.~M. Kristensen
  (Eds.), {STAF} 2020 Workshop Proceedings: 4th Workshop on Model-Driven
  Engineering for the Internet-of-Things, 1st International Workshop on
  Modeling Smart Cities, and 5th International Workshop on Open and Original
  Problems in Software Language Engineering co-located with Software
  Technologies: Applications and Foundations federation of conferences {(STAF}
  2020), Bergen, Norway, June 22-26, 2020, Vol. 2707 of {CEUR} Workshop
  Proceedings, CEUR-WS.org, 2020, pp. 22--34.
\newline\urlprefix\url{http://ceur-ws.org/Vol-2707/mde4iotpaper1.pdf}

\bibitem{Pietron+2021}
J.~Pietron, F.~F\"{u}g, M.~Tichy, An operation-based versioning approach for
  synchronous and asynchronous collaboration in graphical modeling tools, in:
  Proc. of the 1st International Workshop on Foundations and Practice of Visual
  Modeling, Bergen, 21-25 June, 2021, 2021, accepted.

\bibitem{Gackenheimer2015}
C.~Gackenheimer, \href{https://doi.org/10.1007/978-1-4842-1245-5_5}{Introducing
  Flux: An Application Architecture for React}, Apress, Berkeley, CA, 2015, pp.
  87--106.
\newblock \href {http://dx.doi.org/10.1007/978-1-4842-1245-5_5}
  {\path{doi:10.1007/978-1-4842-1245-5_5}}.
\newline\urlprefix\url{https://doi.org/10.1007/978-1-4842-1245-5_5}

\bibitem{Runeson2012}
P.~Runeson, M.~Host, A.~Rainer, B.~Regnell, Case study research in software
  engineering: Guidelines and examples, John Wiley \& Sons, 2012.

\bibitem{Singer2008}
J.~Singer, S.~E. Sim, T.~C. Lethbridge, Software engineering data collection
  for field studies, in: Guide to Advanced Empirical Software Engineering,
  Springer, 2008, pp. 9--34.

\bibitem{Fakih2021}
M.~Fakih, O.~Klemp, S.~Puch, K.~Gr\"{u}ttner, A modeling methodology for
  collaborative evaluation of future automotive innovations, Software and
  Systems Modeling\href
  {http://dx.doi.org/https://doi.org/10.1007/s10270-021-00864-3}
  {\path{doi:https://doi.org/10.1007/s10270-021-00864-3}}.

\end{thebibliography}

\end{document}